\documentclass[
   aps,
   prapplied,
   twocolumn,
   reprint,
   superscriptaddress,
   floatfix,
   citeautoscript,
   longbibliography,
]{revtex4-2}

\usepackage[T1]{fontenc}
\usepackage[utf8]{inputenc}

\usepackage{newtxtext}
\usepackage{newtxmath}

\usepackage{booktabs}
\usepackage{chemformula}
\usepackage{siunitx}
\DeclareSIUnit\atom{atom}
\DeclareSIUnit\torr{Torr}
\DeclareSIUnit\gauss{G}
\DeclareSIUnit\stoppingunit{SU}

\usepackage{graphicx}
\usepackage[
   unicode,
   colorlinks = true,
   allcolors = blue,
]{hyperref}

\usepackage{cleveref}

\usepackage{microtype}

\usepackage{float}

\usepackage{glossaries}
\glsdisablehyper

\newacronym{1d}{1D}{one-dimensional}
\newacronym{2d}{2D}{two-dimensional}
\newacronym{3d}{3D}{three-dimensional}

\newacronym{ac}{AC}{alternating current}
\newacronym{afm}{AFM}{atomic force microscopy}
\newacronym{alc}{ALC}{avoided level crossing}
\newacronym{api}{API}{application programming interface}
\newacronym{ariel}{ARIEL}{Advanced Rare Isotope Laboratory}
\newacronym{arpes}{ARPES}{angle-resolved photoemission spectroscopy}
\newacronym{atp}{ATP}{adenosine triphosphate}

\newacronym[sort={b-NMR}]{bnmr}{\ensuremath{\beta}-NMR}{\ensuremath{\beta}-detected nuclear magnetic resonance}
\newacronym[sort={b-NQR}]{bnqr}{\ensuremath{\beta}-NQR}{\ensuremath{\beta}-detected nuclear quadrupole resonance}
\newacronym{bca}{BCA}{binary collision approximation}
\newacronym{bcc}{BCC}{body-centred cubic}
\newacronym{bcp}{BCP}{buffered chemical polishing}
\newacronym{bcs}{BCS}{Bardeen-Cooper-Schrieffer}
\newacronym{bpp}{BPP}{Bloembergen-Purcell-Pound}
\newacronym{bsc}{BSC}{\ch{Bi2Se3:Ca}}
\newacronym{btm}{BTM}{\ch{Bi2Te3:Mn}}
\newacronym{bts}{BTS}{\ch{Bi2Te2Se}}

\newacronym{camp}{CAMP}{control and monitor program}
\newacronym{ccd}{CCD}{charge-coupled device}
\newacronym{cdw}{CDW}{charge density wave}
\newacronym{cgs}{CGS}{centimetre-gram-second system of units}
\newacronym{cmms}{CMMS}{Centre for Molecular and Materials Science}
\newacronym{codata}{CODATA}{Committee on Data for Science and Technology}
\newacronym{cpu}{CPU}{central processing unit}
\newacronym{create}{CREATE}{Collaborative Research and Training Experience Program}
\newacronym{cw}{CW}{continuous wave}

\newacronym{daq}{DAQ}{data acquisition}
\newacronym{dc}{DC}{direct current}
\newacronym{dft}{DFT}{density functional theory}
\newacronym{dos}{DOS}{density of states}
\newacronym{dqt}{DQT}{double-quantum transition}

\newacronym{efg}{EFG}{electric field gradient}
\newacronym{emim-ac}{EMIM-Ac}{1-ethyl-3-methylimidazolium acetate}
\newacronym{emim-dca}{EMIM-DCA}{1-ethyl-3-methylimidazolium dicyanamide}
\newacronym{epr}{EPR}{electron paramagnetic resonance}
\newacronym{esr}{EPR}{electron spin resonance}
\newacronym{endor}{ENDOR}{electron nuclear double resonance}
\newacronym{epics}{EPICS}{Experimental Physics and Industrial Control System}

\newacronym{fcc}{FCC}{face-centred cubic}
\newacronym{fft}{FFT}{fast Fourier transform}
\newacronym{fom}{FoM}{figure of merit}
\newacronym{fwhm}{FWHM}{full width at half maximum}

\newacronym{gga}{GGA}{generalized gradient approximation}

\newacronym{hb}{HB}{hole-burning}
\newacronym{hfqs}{HFQS}{high-field \ensuremath{Q} slope}
\newacronym{hv}{HV}{high-voltage}
\newacronym{hwhm}{HWHM}{half width at half maximum}

\newacronym{iaea}{IAEA}{International Atomic Energy Agency}
\newacronym{il}{IL}{ionic liquid}
\newacronym{is}{IS}{impedance spectroscopy}
\newacronym{isac}{ISAC}{isotope separator and accelerator}
\newacronym{isol}{ISOL}{isotope separation online}
\newacronym{isosim}{IsoSiM}{Isotopes for Science and Medicine}

\newacronym{lcao}{LCAO}{linear combination of atomic orbitals}
\newacronym{lda}{LDA}{local density approximation}
\newacronym{led}{LED}{light-emitting diode}
\newacronym{leis}{LEIS}{low-energy ion scattering}
\newacronym{lib}{LIB}{lithium-ion battery}
\newacronym{lsat}{LSAT}{\ch{(La,Sr)(Al,Ta)O3}}

\newacronym{mas}{MAS}{magic angle spinning}
\newacronym{mpms}{MPMS}{magnetic property measurement system}
\newacronym{mbe}{MBE}{molecular beam epitaxy}
\newacronym{md}{MD}{molecular dynamics}
\newacronym{midas}{MIDAS}{Maximum Integrated Data Acquisition System}
\newacronym{mit}{MIT}{metal-insulator transition}
\newacronym{mnr}{MNR}{Meyer-Neldel rule}
\newacronym{mqt}{mqt}{multi-quantum transition}
\newacronym{mud}{MUD}{muon data}
\newacronym{ms}{MS}{mass spectrometry}

\newacronym{nbm}{NBM}{neutral beam monitor}
\newacronym{neb}{NEB}{nudged elastic band}
\newacronym{nim}{NIM}{nuclear instrumentation module}
\newacronym{nmr}{NMR}{nuclear magnetic resonance}
\newacronym{no}{NO}{nuclear orientation}
\newacronym{nqr}{NQR}{nuclear quadrupole resonance}
\newacronym{nrc}{NRC}{National Research Council of Canada}
\newacronym{nserc}{NSERC}{Natural Sciences and Engineering Research Council of Canada}

\newacronym{oa}{OA}{optical absorption}

\newacronym{pac}{PAC}{perturbed angular correlation}
\newacronym{pad}{PAD}{perturbed angular distribution}
\newacronym{pas}{PAS}{principle axis system}
\newacronym{pchip}{PCHIP}{piecewise cubic Hermite interpolating polynomial}
\newacronym{pdf}{PDF}{probability density function}
\newacronym{pld}{PLD}{pulsed laser deposition}
\newacronym{ppms}{PPMS}{physical property measurement system}

\newacronym{qens}{QENS}{quasielastic neutron scattering}
\newacronym{ql}{QL}{quintuple layer}
\newacronym{qo}{QO}{quantum oscillations}

\newacronym{rbs}{RBS}{Rutherford backscattering}
\newacronym{rf}{RF}{radio frequency}
\newacronym{rheed}{RHEED}{reflection high-energy electron diffraction}
\newacronym{rib}{RIB}{radioactive ion beam}
\newacronym{rkky}{RKKY}{Ruderman–Kittel–Kasuya–Yosida}
\newacronym{rrr}{RRR}{residual-resistivity ratio}
\newacronym{rtil}{RTIL}{room temperature ionic liquid}

\newacronym{sae}{SAE}{spin-alignment echo}
\newacronym{sans}{SANS}{small angle neutron scattering}
\newacronym{si}{SI}{International System of Units}
\newacronym{sims}{SIMS}{secondary ion mass spectrometry}
\newacronym{slr}{SLR}{spin-lattice relaxation}
\newacronym[sort={S/N}]{snr}{\textit{S}/\textit{N}}{signal-to-noise ratio}
\newacronym{squid}{SQUID}{superconducting quantum interference device}
\newacronym{srf}{SRF}{superconducting radio frequency}
\newacronym{srim}{SRIM}{Stopping and Range of Ions in Matter}
\newacronym{ssid}{SSID}{solid-state ionic device}
\newacronym{ssr}{SSR}{spin-spin relaxation}
\newacronym{stm}{STM}{scanning tunnelling microscopy}
\newacronym{sts}{STS}{scanning tunnelling spectroscopy}

\newacronym{ti}{TI}{topological insulator}
\newacronym{trim}{TRIM}{Transport and Range of Ions in Matter}
\newacronym{tss}{TSS}{topological surface state}
\newacronym{tmd}{TMD}{transition metal dichalcogenide}

\newacronym{uhv}{UHV}{ultra-high vacuum}

\newacronym{vdw}{vdW}{van der Waals}
\newacronym{vft}{VFT}{Vogel-Fulcher-Tammann}

\newacronym{xrd}{XRD}{x-ray diffraction}
\newacronym{xrr}{XRR}{x-ray reflection}

\newacronym{ybco}{YBCO}{\ch{YBa2Cu3O_{6+x}}}
\newacronym{ysz}{YSZ}{yttria-stabilized zirconia}

\newacronym[sort={muSR}]{musr}{\ensuremath{\mu}SR}{muon spin rotation/relaxation/resonance}
\newacronym{alc-musr}{ALC-\ensuremath{\mu}SR}{avoided level crossing muon spin rotation}
\newacronym{le-musr}{LE-\ensuremath{\mu}SR}{low-energy muon spin rotation}
\newacronym{lf-musr}{LF-\ensuremath{\mu}SR}{longitudinal field muon spin rotation}
\newacronym{rf-musr}{RF-\ensuremath{\mu}SR}{radio frequency muon spin rotation}
\newacronym{tf-musr}{TF-\ensuremath{\mu}SR}{transverse field muon spin rotation}
\newacronym{zf-musr}{ZF-\ensuremath{\mu}SR}{zero field muon spin rotation}

\DeclareMathOperator{\erf}{erf}
\DeclareMathOperator{\erfi}{Erfi}

\begin{document}

\title{
	Depth-resolved measurements of the Meissner screening profile in surface-treated \ch{Nb}
}

% coauthors
\author{Ryan~M.~L.~McFadden}
\email[E-mail: ]{rmlm@triumf.ca}
\affiliation{TRIUMF, 4004 Wesbrook Mall, Vancouver, BC V6T~2A3, Canada}
\affiliation{Department of Physics and Astronomy, University of Victoria, 3800 Finnerty Road, Victoria, BC V8P~5C2, Canada}

\author{Md~Asaduzzaman}
\affiliation{TRIUMF, 4004 Wesbrook Mall, Vancouver, BC V6T~2A3, Canada}
\affiliation{Department of Physics and Astronomy, University of Victoria, 3800 Finnerty Road, Victoria, BC V8P~5C2, Canada}

\author{Thomas~Prokscha}
\affiliation{Laboratory for Muon Spin Spectroscopy, Paul Scherrer Institute, Forschungsstrasse 111, 5232 Villigen, Switzerland}

\author{Zaher~Salman}
\affiliation{Laboratory for Muon Spin Spectroscopy, Paul Scherrer Institute, Forschungsstrasse 111, 5232 Villigen, Switzerland}

\author{Andreas~Suter}
\affiliation{Laboratory for Muon Spin Spectroscopy, Paul Scherrer Institute, Forschungsstrasse 111, 5232 Villigen, Switzerland}

\author{Tobias~Junginger}
\email[E-mail: ]{junginger@uvic.ca}
\affiliation{TRIUMF, 4004 Wesbrook Mall, Vancouver, BC V6T~2A3, Canada}
\affiliation{Department of Physics and Astronomy, University of Victoria, 3800 Finnerty Road, Victoria, BC V8P~5C2, Canada}

% today's date
\date{\today}

\begin{abstract}
	We report depth-resolved measurements of the Meissner screening
	profile in several surface-treated \ch{Nb} samples using \gls{le-musr}.
	In these experiments,
	implanted positive muons,
	whose stopping depths below \ch{Nb}'s surface were adjusted between \SIrange{\sim 10}{\sim 150}{\nano\meter},
	reveal the field distribution inside the superconducting element via their spin-precession
	(communicated through their radioactive decay products).
	We compare how the field screening is modified by different surface treatments
	commonly employed to prepare \gls{srf} cavities used in accelerator beamlines.
	In contrast to an earlier report
	[A.~Romanenko \emph{et al.}, \href{https://doi.org/10.1063/1.4866013}{Appl.\ Phys.\ Lett.\ \textbf{104} 072601 (2014)}],
	we find no evidence for any ``anomalous'' modifications to the Meissner profiles,
	with all data being well-described by a London model.
	Differences in screening properties between surface treatments can be
	explained by changes to the carrier mean-free-paths resulting from 
	dopant profiles near the material's surface.
\end{abstract}

\maketitle
\glsresetall

\section{
	Introduction
	\label{sec:introduction}
}

Since its discovery in the early 19\textsuperscript{th} century,
the elemental metal \ch{Nb} has been the subject of intense study,
culminating in a detailed understanding of its chemical properties
(see e.g.,~\cite{1999-Nowak-CR-99-3603,2009-Schlewitz-KOECT-17-1,2017-Arblaster-JPED-38-707}).
Of equal (if not greater) interest are the element's electronic properties,
especially those relating to its superconductivity.
For example,
while \ch{Nb} is one of the many superconducting elements~\cite{2004-Buzea-SST-18-R1},
its transition temperature $T_{c} \approx \SI{9.25}{\kelvin}$
is the highest among them at ambient pressure.
Accompanying this accolade is a lower critical field $B_{c1} \approx \SI{170}{\milli\tesla}$
that exceeds all other (type-II) superconductors,
making the element particularly suited for devices that must remain flux-free
under modest magnetic fields.
These intrigues inspired comprehensive measurements~\cite{1964-Leupold-PR-134-A1322,1969-Webb-PR-181-1127,1978-Karim-JLTP-30-389,2000-Chainani-PRL-85-1966}
and calculations~\cite{1970-Mattheiss-PRB-1-373,1979-Crabtree-PRL-42-390,1981-Pinski-PRB-23-5080,1984-Blaschke-JPFMP-14-175,1987-Crabtree-PRB-35-1728,1991-Weber-PRB-44-7585}
relating to its electronic structure,
as well as how it is modified under pressure~\cite{1983-Neve-PRB-28-629}.
This in turn has led to a thorough understanding of its
superconductivity~\cite{1965-Maxfield-PR-139-A1515,1966-Finnemore-PR-149-231,1968-French-C-8-301},
including details
such as its ``strong''~\cite{1967-Nam-PR-156-470} electron-phonon coupling~\cite{1996-Savrasov-PRB-54-16487,1998-Bauer-PRB-57-11276,2020-Giri-MTP-12-100175},
non-local electrodynamics~\cite{1953-Pippard-PRSLA-216-547,1957-Bardeen-PR-108-1175} in ``clean'' samples~\cite{2005-Suter-PRB-72-024506},
and
vortex lattice structure~\cite{2013-Maisuradze-PRB-88-140509,2014-Yaouanc-PRB-89-184503,2015-Reimann-NC-6-8813}.
In fact,
this high degree of understanding has prompted \ch{Nb}'s frequent use complex heterostructures where
a superconducting layer is required
(see e.g.,~\cite{2014-Flokstra-PRB-89-054510,2015-DiBernardo-PRX-5-041021,2016-Flokstra-NP-12-57,2018-Flokstra-PRL-120-247001,2019-Flokstra-APL-115-072602,2019-Stewart-PRB-100-020505,2020-Krieger-PRL-125-026802,2021-Rogers-CP-4-69,2021-Flokstra-RPB-104-L060506,2021-Alpern-PRM-5-114801}).
These physical traits,
in conjunction with the metal's mechanical properties~\cite{2007-Myneni-AIPCP-927-41,2015-Ciovati-MSEA-642-117},
have also made \ch{Nb} particularly suited for use in
\gls{srf} cavities~\cite{2008-Padamsee-RFSA-2,2009-Padamsee-RFSSTA,2017-Padamsee-SST-30-053003},
which use a large electric field, $E_{\mathrm{acc}}$, to accelerate charged
particles in accelerator beamlines.
Of particular importance for this application
is the maximum achievable value for $E_{\mathrm{acc}}$,
which is fundamentally limited by \ch{Nb}'s ability to expel magnetic flux
(i.e., to remain in the Meissner state).
This application, in particular,
has been a driving factor in the continued refinement
of our understanding of the element.

Central to \ch{Nb}'s performance in \gls{srf} cavities is the preparation of its surface,
with many empirical ``recipes'' developed
(e.g.,
low-temperature baking~\cite{2004-Ciovati-JAP-96-1591},
two-step baking~\cite{arXiv:1806.09824},
nitrogen doping~\cite{2013-Grassellino-SST-26-102001},
nitrogen infusion~\cite{2017-Grassellino-SST-30-094004},
etc.)
explicitly for boosting cavity performance
(i.e., maximizing its quality factor $Q$ for the largest possible range of $E_{\mathrm{acc}}$).
While it has long been known that surfaces play an important role
in determining the field of first-flux-entry
(see e.g.,~\cite{1964-Bean-PRL-12-14}),
some significant progress has been made recently.
For example,
it has been demonstrated that \ch{Nb} can expel flux up to its so-called
superheating field $B_{\mathrm{sh}}$
(see e.g.,~\cite{2011-Transtrum-PRB-83-094505})
by coating it with a thin superconducting film~\cite{2017-Junginger-SST-30-125012},
with sample geometry and surface preparations also playing an important role~\cite{2018-Junginger-PRAB-21-032002}.
For the latter,
the effect can be subtle for samples with identical geometries~\cite{2022-Turner-SR-12-5522};
however,
measurements of their Meissner screening profiles revealed significant differences
between select treatments~\cite{2014-Romanenko-APL-104-072601},
with low-temperature baking~\cite{2004-Ciovati-JAP-96-1591}
postulated at creating an ``effective''
superconductor-superconductor bilayer~\cite{2014-Kubo-APL-104-032603,2017-Kubo-SST-30-023001,2019-Kubo-JJAP-58-088001}
near the surface
(i.e., from a thin region of ``dirty'' \ch{Nb} near the surface on top of the ``clean'' bulk).
While this has stimulated a renewed interesting in using multilayers for \gls{srf} applications,
the results are controversial and warrant further investigation.

Currently, there are relatively few experimental techniques with the right
combination of electromagnetic and spatial sensitivity to achieve this goal
(e.g., ion-implanted \gls{bnmr}~\cite{2015-MacFarlane-SSNMR-68-1,2022-MacFarlane-ZPC-236-757}
and
\gls{le-musr}~\cite{2004-Bakule-CP-45-203,2004-Morenzoni-JPCM-16-S4583,2022-Hillier-NRMP-2-4}).
One possibility is to use \gls{le-musr}~\cite{2004-Bakule-CP-45-203,2022-Hillier-NRMP-2-4},
which uses muons implanted in the near-surface region
(depths less than \SI{\sim 150}{\nano\meter}) as local ``magnetometers''.
This technique has been used to study Meissner screening in \ch{Nb} with success,
revealing:
the importance of non-local electrodynamics~\cite{1953-Pippard-PRSLA-216-547,1957-Bardeen-PR-108-1175}
and strong-coupling corrections~\cite{1967-Nam-PR-156-470} in ``clean'' samples~\cite{2005-Suter-PRB-72-024506};
the impact of growth methods on the screening properties of \ch{Nb/Cu} films used in \gls{srf} cavities~\cite{2017-Junginger-SST-30-125013};
as well as the aforementioned ``anomalous'' modification in the character of the screening profile
by mild baking~\cite{2014-Romanenko-APL-104-072601}.
In order to clarify the origin of the latter,
here we used \gls{le-musr} to quantify the screening profile in \ch{Nb}
samples with surface treatments commonly employed in \gls{srf}
applications~\cite{2004-Ciovati-JAP-96-1591,arXiv:1806.09824,2017-Grassellino-SST-30-094004}.
Our results revealed no ``anomalous'' modifications to the screening profiles for \emph{any} of the surface treatments,
with all profiles being well-described by a London model~\cite{1935-London-PRSLA-149-71}.
The surface treatments were found to produce different magnetic penetrations depths,
which can be explained by dissimilar carrier mean-free-paths within the first \SI{\sim 150}{\nano\meter}
below the surface.

\section{
	Experiment
	\label{sec:experiment}
}

\begin{figure}
	\centering
	\includegraphics[width=1.0\columnwidth]{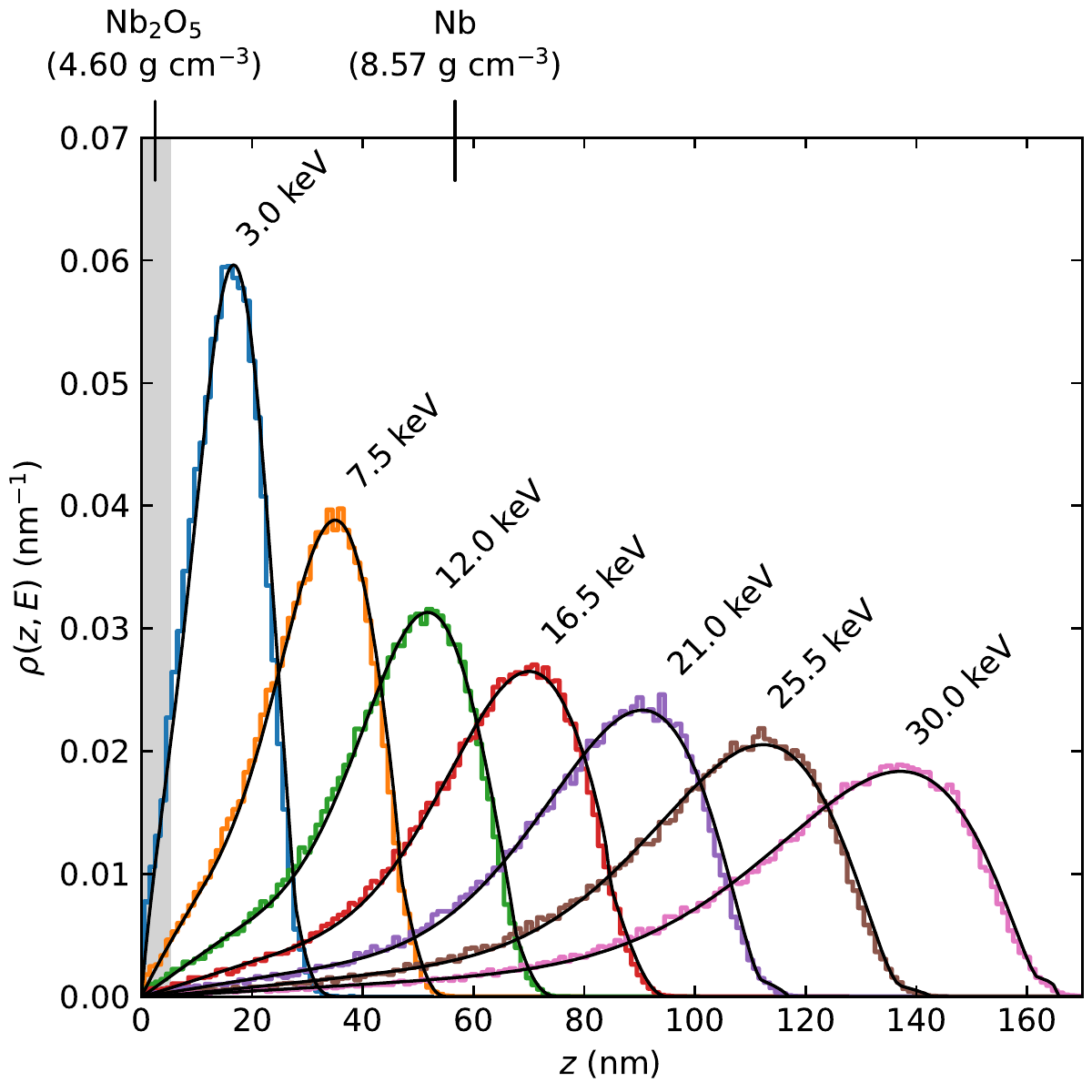}
	\caption{
		\label{fig:implantation-profiles}
		Typical stopping profiles $\rho(z, E)$ for $\mu^{+}$ implanted in a \ch{Nb2O5}(\SI{5}{\nano\meter})/\ch{Nb} target
		at different energies $E$ (indicated in the inset),
		simulated using the Monte Carlo code
		TRIM.SP~\cite{1984-Eckstein-NIMB-2-550,1991-Eckstein-SSMS-10,1994-Eckstein-REDS-1-239}.
		The profiles, represented as histograms with \SI{1}{\nano\meter} bins,
		were generated from \num{e5} projectiles.
		The solid black lines denote fits to a model for $\rho(z, E)$
		[\Cref{eq:stopping,eq:beta-pdf} --- see \Cref{sec:results}],
		clearly capturing all features of the individual profiles.
		Additional simulation details can be found in \Cref{sec:trimsp}.
	}
\end{figure}

\Gls{le-musr} experiments were performed at the Paul Sherrer Institute's
Swiss Muon Source (located in Villigen, Switzerland).
Using the $\mu E4$ beamline~\cite{2008-Prokscha-NIMA-595-317},
``low-energy'' muons were generated by moderating the energy of a
\SI{\sim 4}{\mega\electronvolt}  ``surface'' muon beam
using a film of condensed cryogenic gas~\cite{1994-Morenzoni-PRL-72-2793,2001-Prokscha-ASS-172-235}
and electrostatically re-accelerating the eluting epithermal (\SI{\sim 15}{\electronvolt}) muons
to energies on the order of \SI{\sim 15}{\kilo\electronvolt}.
The resulting beam,
with a typical intensity of \SI{\sim e4}{\per\second},
was delivered to a dedicated spectrometer~\cite{2000-Morenzoni-PB-289-653,2008-Prokscha-NIMA-595-317,2012-Salman-PP-30-55}
using electrostatic optics housed within an \gls{uhv} beamline.
The $\mu^{+}$ arrival times were triggered on a thin
(\SI{\sim 10}{\nano\meter}) carbon foil detector,
causing a slight reduction in their mean kinetic energy
(\SI{\sim 1}{\kilo\electronvolt})
and an (asymmetric) energy spread (\SI{\sim 450}{\electronvolt})
before reaching the sample.
Control over the $\mu^{+}$ implantation energy was achieved by biasing
an electrically isolated sample holder using a \gls{hv} power supply,
providing access to stopping depths between \SIrange{\sim 10}{\sim 150}{\nano\meter}
below the sample surface.
The stopping of $\mu^{+}$ in solids can be accurately computed~\cite{2002-Morenzoni-NIMB-192-245}
using Monte Carlo codes
(e.g., TRIM.SP~\cite{1984-Eckstein-NIMB-2-550,1991-Eckstein-SSMS-10,1994-Eckstein-REDS-1-239}),
which we used here to simulate $\mu^{+}$ stopping profiles in \ch{Nb}
(see \Cref{fig:implantation-profiles}).
To ensure the accuracy of these predictions,
we revised the parameterization of \ch{Nb}'s electronic stopping cross section
for proton-like projectiles~\cite{1977-Anderson-SRIM-3,1993-ICRU-49}
using a Varelas-Biersack fit~\cite{1970-Varelas-NIM-79-213}
to an up-to-date compilation~\cite{2017-Montanari-NIMB-408-50}
of experimental values~\cite{1984-Sirotinin-NIMB-4-337, 1986-Bauer-NIMB-13-201, 1988-Ogino-NIMB-33-155, 2020-Moro-PRA-102-022808}. 
The revised fit suggests that earlier calculations 
(see e.g.,~\cite{2005-Suter-PRB-72-024506,2014-Flokstra-PRB-89-054510,2014-Romanenko-APL-104-072601,2015-DiBernardo-PRX-5-041021,2016-Flokstra-NP-12-57,2017-Junginger-SST-30-125013,2018-Flokstra-PRL-120-247001,2019-Flokstra-APL-115-072602,2019-Stewart-PRB-100-020505,2020-Krieger-PRL-125-026802,2021-Rogers-CP-4-69,2021-Flokstra-RPB-104-L060506,2021-Alpern-PRM-5-114801})
likely underestimate the $\mu^{+}$ range in \ch{Nb}
(or \ch{Nb} layers).
Further details are given in \Cref{sec:trimsp}.

The basis of the \gls{le-musr} technique~\cite{2004-Bakule-CP-45-203,2004-Morenzoni-JPCM-16-S4583,2021-Prokscha-MSI-18-274,2022-Hillier-NRMP-2-4}
involves implanting a beam of (\SI{\sim 100}{\percent}) spin-polarized $\mu^{+}$
into a sample of interest and observing their spins, $\mathbf{S}$,
reorient in their local magnetic field, $\mathbf{B}$.
This process is monitored via the anisotropic $\beta$-emissions
from $\mu^{+}$ decay
(mean lifetime $\tau_{\mu} = \SI{2.1969811 \pm 0.0000022}{\micro\second}$~\cite{2020-Zyla-PTEP-2020-083C01}),
wherein the direction of the emitted $\beta$-rays are probabilisitically correlated
with $\mathbf{S}$ at the moment of decay
(see e.g.,~\cite{2011-Yaouanc-MSR}).
When $\mathbf{B}$ is transverse to the spin direction,
$\langle S \rangle$ will precess at a rate equal to the probe's Larmor frequency:
\begin{equation}
	\omega_{\mu} = \gamma_{\mu} B ,
\end{equation}
where $\gamma_{\mu} / (2 \pi) = \SI{135.538 809 4 \pm 0.000 003 0}{\mega\hertz\per\tesla}$
is the muon gyromagnetic ratio~\cite{2021-Tiesinga-RMP-93-025010}.
In the experiments performed here,
this so-called transverse-field geometry was used
(see e.g.,~\cite{2004-Bakule-CP-45-203,2011-Yaouanc-MSR}),
wherein an external field $B_\mathrm{applied} \approx \SI{25}{\milli\tesla}$
was applied perpendicular to the initial direction of $\mu^{+}$ spin-polarization
and parallel to the surface of our \ch{Nb} samples.
This configuration is highly sensitive to inhomogeneities in $B$
(as expected near the surface of a superconductor in the Meissner state),
allowing for the local field distribution, $p(B)$, to be measured,
which reflects the the screening properties of the samples.

In our \gls{le-musr} measurements,
the temporal evolution of the \emph{asymmetry}, $A(t)$,
in the $\beta$-rates recorded for two opposing detectors
(i.e., \SI{180}{\degree} opposite one another),
was monitored after $\mu^{+}$ implantation.
The counts in a single detector, $N_{\pm}$, are given by:
\begin{equation}
	\label{eq:counts}
	N_{\pm}(t) = N_{0, \pm} \exp \left ( - \frac{t}{\tau_{\mu}} \right ) \left [ 1 \pm A(t) \right ] + b_{\pm} ,
\end{equation}
where $N_{0, \pm}$ and $b_{\pm}$ are denote the incoming rates of
``good'' and ``background'' decay events.
While \Cref{eq:counts} can be re-arranged for $A(t)$,
in a two-counter experiment it can be determined directly from the normalized difference
in $\beta$-rates from the two counters:
\begin{equation}
  	\label{eq:asymmetry}
  	A(t) \equiv \frac{ \left [ N_{+}(t) - b_{+} \right ] - \alpha \left [ N_{-}(t) - b_{-} \right ] }{ \left [ N_{+}(t) - b_{+} \right ] + \alpha \left [ N_{-}(t) - b_{-} \right ] } = A_{0} P_{\mu}(t) ,
\end{equation}
where $A_{0}$ is a constant whose precise value depends on both the geometry of the experiment
and the details of $\mu^{+}$ decay,
and the factor $\alpha \equiv N_{0, +} / N_{0, -}$
accounts for differences between the detector pair
(e.g., detection efficiencies, solid angle coverage, etc.~\cite{1994-Riseman-HI-87-1135,2011-Yaouanc-MSR}).
The time-dependence in \Cref{eq:asymmetry} is determined entirely
by the spin-polarization of the muon ensemble, $P_{\mu}(t)$,
which depends on $p(B)$ according to:
\begin{equation}
	\label{eq:polarization}
	P_{\mu}(t) = \int_{-\infty}^{+\infty} p(B) \cos ( \omega_{\mu} t + \phi ) \, \mathrm{d}B ,
\end{equation}
where $t$ is the time (in \si{\micro\second}) after implantation
and
$\phi$ is a phase factor that depends on the experimental setup
(approximately \SI{-40}{\degree} here).
Note that \Cref{eq:polarization} makes the simplifying assumption that $P_{\mu}(0) \approx 1$
(i.e., the $\mu^{+}$ are initially \SI{\sim 100}{\percent} spin-polarized).
Thus,
from the synergistic information encapsulated within $P_{\mu}(t)$
and the simulated $\mu^{+}$ stopping profiles
(see \Cref{fig:implantation-profiles}),
it is feasible to reconstruct how $B$ varies with depth, $z$,
below the sample surface
(see below).

\subsection{
	Samples
	\label{sec:experiment:samples}
}

In accord with the standard practice used when fabricating \gls{srf} cavities~\cite{2008-Padamsee-RFSA-2,2009-Padamsee-RFSSTA,2017-Padamsee-SST-30-053003},
all samples were sourced from high \gls{rrr} \ch{Nb}
(i.e., $\mathrm{RRR} \gtrsim 300$).
Each sample consisted of a piece of the ``stock'' metal
machined to into a flat plate
(\SI{\sim 25 x \sim 25 x \sim 1.5}{\milli\meter})
with a small circular aperture
(\SI{\sim 6.5}{\milli\meter} diameter)
in one corner.
The pieces were then hand polished to remove any sharp edges,
followed by \gls{bcp} to remove any damaged layers near the surface
(see e.g.,~\cite{2011-Ciovati-JAE-41-721}).
Subsequently, the samples were annealed at \SI{1400}{\celsius} for \SI{5}{\hour}
to remove any mechanical stresses remaining in the metal.
Afterwards, an additional round of \gls{bcp} was performed to
remove the topmost \SI{\sim 10}{\micro\meter} of material from the surface
(i.e., to remove any contaminants introduced from the oven during annealing).
In the remainder of this manuscript,
we denote this as the ``baseline'' surface treatment for \gls{srf} cavity grade \ch{Nb}.
This process has been shown to remove virtually all pinning~\cite{2018-Junginger-PRAB-21-032002}.

On top of the ``baseline'' preparation,
several samples underwent additional surface treatments.
One sample was baked at \SI{120}{\celsius} for \SI{48}{\hour} in \gls{uhv}~\cite{2004-Ciovati-JAP-96-1591},
which we call ``\SI{120}{\celsius} bake''.
Another sample underwent a two-step baking procedure,
wherein it was initially heated to \SI{75}{\celsius} for \SI{5}{\hour} and
then additionally to \SI{120}{\celsius} for \SI{48}{\hour}~\cite{arXiv:1806.09824},
which we denote as ``\SI{75}{\celsius}/\SI{120}{\celsius} bake''.
Lastly,
one sample was initially heated to \SI{800}{\celsius} for \SI{3}{\hour} under high vacuum
and subsequently baked at \SI{120}{\celsius} for \SI{48}{\hour} in a \SI{25}{\milli\torr} \ch{N_{2}} atmosphere~\cite{2017-Grassellino-SST-30-094004},
which we refer to as ``\ch{N2} infusion''.
A magnetometric characterization of \ch{Nb} samples with identical surface treatments 
can be found elsewhere~\cite{2022-Turner-SR-12-5522}.

\begin{figure}
	\centering
	\includegraphics[width=1.0\columnwidth]{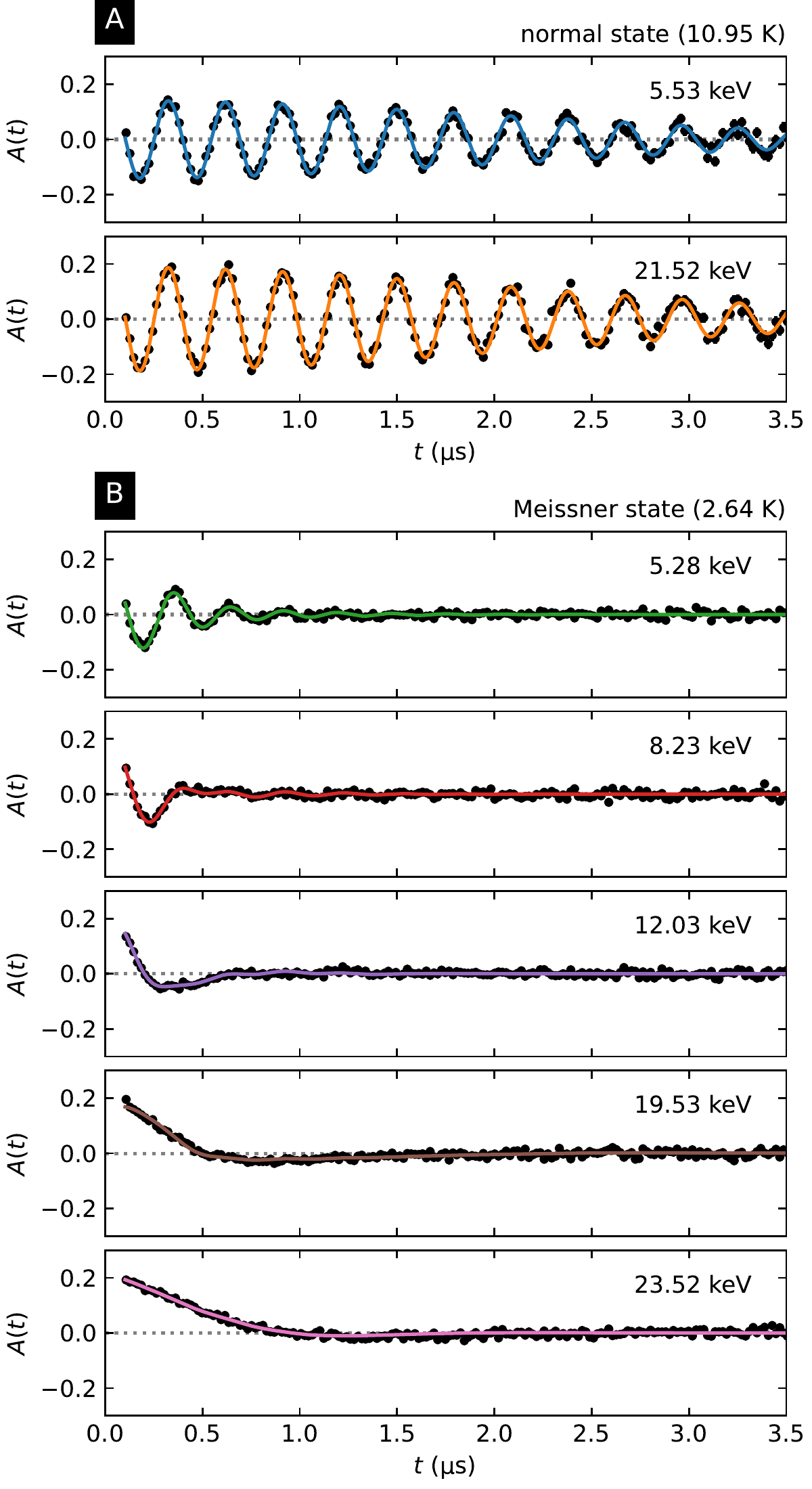}
	\caption{
		\label{fig:baseline-spectra}
		Implantation energy dependence of transverse-field \gls{le-musr} data
		in \ch{Nb} (``baseline''),
		measured in both the normal (\SI{\sim 11}{\kelvin})
		and Meissner (\SI{\sim 2.6}{\kelvin}) states
		with an applied magnetic field of \SI{\sim 25}{\milli\tesla}.
		The $\mu^{+}$ energy $E$ is indicated in the inset of each subplot.
		In the normal state
		(\textbf{A})
		there is no significant energy dependence
		to the temporal evolution of $A(t)$,
		implying that all implanted muons sample the same local field
		distribution $p(B)$ below the sample surface.
		By contrast, in Meissner state
		(\textbf{B})
		$A(t)$ depends strongly on the implantation energy.
		As the implantation energy increases,
		the $\mu^{+}$ spin-precession frequency decreases,
		accompanied by increased damping of the signal,
		consistent with screening of the magnetic field with increasing depths
		below the sample surface.
		The coloured lines denote a fit to \emph{all} of the the data
		(i.e., a global fit) using
		\Cref{eq:fit-function,eq:polarization-skewed-gaussian,eq:polarization-skewed-gaussian-parts,eq:polarization-gaussian},
		where the phase $\phi$ was shared as a common parameter.
		Clearly, the model captures all of the data's main features.
		Note that the Gaussian term in \Cref{eq:fit-function} accounts for
		the small (\SI{< 10}{\percent}) fraction of muons that do not stop in the sample
		(e.g., due to backscattering).
	}
\end{figure}

\section{
	Results \& Analysis
	\label{sec:results}
}

Typical time-differential \gls{le-musr} data for our surface-treated \ch{Nb} samples
are shown in \Cref{fig:baseline-spectra}.
In the normal state ($T > T_{c}$),
there is no significant energy dependence to the temporal evolution of $A(t)$,
indicating that all implanted muons sample the same local field
distribution below the sample surface.
This is evident from the identical precession frequencies
and damping envelopes,
the latter being (predominantly) a result of the host \ch{^{93}Nb} nuclei
(spin $I = 9/2$;
$\gamma/(2\pi) = \SI{10.4523 \pm 0.0005}{\mega\hertz\per\tesla}$;
\SI{100}{\percent} natural abundance)~\cite{2011-Baglin-NDS-112-1163}.
By contrast,
$A(t)$ depends strongly on implantation energy in Meissner state.
As the implantation energy increases,
the $\mu^{+}$ spin-precession frequency decreases,
accompanied by substantial damping of the signal.
These features are expected for a broad $p(B)$
whose mean shifts to lower values at increasing depths below the surface,
consistent with the expected ``signature'' for screening of the applied magnetic field.

To quantify these details,
we now consider an analysis of the data,
which amounts to choosing an (analytic) approximation
for the field distribution in \Cref{eq:polarization}.
Often, $p(B)$ can be approximated by a Gaussian distribution
(see e.g,~\cite{2011-Yaouanc-MSR}):
\begin{equation}
	\label{eq:gaussian}
	p_{\mathrm{G}}(B) = \frac{1}{\sqrt{2 \pi} } \left ( \frac{\gamma_{\mu}}{\sigma} \right ) \exp \left \{ -\frac{1}{2} \left [ \frac{B - B_{0}}{ \left ( \sigma / \gamma_{\mu} \right ) } \right ]^{2} \right \} ,
\end{equation}
where $B_{0}$ and $\sigma$ denote the distribution's location (i.e., mean) and width, respectively.
Upon substitution of \Cref{eq:gaussian} for $p(B)$ into \Cref{eq:polarization},
one gets:
\begin{equation}
	\label{eq:polarization-gaussian}
	P_{\mathrm{G}} = \exp \left ( -\frac{\sigma^{2} t^{2}}{2} \right ) \cos \left ( \gamma_{\mu} B_{0} t + \phi \right ) ,
\end{equation}
which says that the observed precession frequency is given by
the mean of the distribution and that the degree of damping is determined by its width.
While this symmetric distribution works well in many instances,
the field distribution below the surface of a material in the Meissner state
is expected to be intrinsically \emph{asymmetric}
(i.e., because the applied field decays to zero inside the material).
Therefore,
a better approximation for $p(B)$ in our samples
is given by a \emph{skewed} Gaussian~\cite{2008-Suter-M}:
\begin{widetext}
\begin{equation}
	\label{eq:skewed-gaussian}
	p_{\mathrm{SG}}(B) = \sqrt{\frac{2}{\pi}} \left ( \frac{ \gamma_{\mu} }{ \sigma_{-} + \sigma_{+} } \right ) \times \begin{cases}
		\displaystyle \exp \left \{ - \frac{1}{2} \left [ \frac{B - B_{0} }{ \left ( \sigma_{-} / \gamma_{\mu} \right ) } \right ]^{2} \right \} , & \text{for } B < B_{0} , \\
		1, & \text{for } B = B_{0}, \\
		\displaystyle \exp \left \{ - \frac{1}{2} \left [ \frac{B - B_{0} }{ \left ( \sigma_{+} / \gamma_{\mu} \right ) } \right ]^{2} \right \} , & \text{for } B > B_{0} , 
	\end{cases}
\end{equation}
\end{widetext}
where $B_{0}$ is the ``peak'' field
(i.e., \emph{not} the mean of the distribution)
and
$\sigma_{\pm}$ define the distribution's width
(i.e., on either side of $B_{0}$).
Note that the definition in \Cref{eq:skewed-gaussian} is somewhat unusual for a
skewed Gaussian distribution;
it is more commonly defined as:
\begin{equation*}
	% \label{eq:skewed-gaussian-conventional}
	p_{\mathrm{SG}}(B) = p_{\mathrm{G}}(B) \left ( 1  + \erf \left \{ \frac{\varsigma}{\sqrt{2}} \left [ \frac{ B - B_{0}} { \left ( \sigma / \gamma_{\mu} \right )} \right ]  \right \} \right ),
\end{equation*}
where $\erf (z)$ is the error function and $\varsigma \in [-\infty, +\infty]$
is the ``skewness'' parameter~\footnote{In contrast to the ``unusual'' definition in \Cref{eq:skewed-gaussian}, the ``conventional'' expression for $p_{\mathrm{SG}}(B)$ relies on ``weighting'' $p_{\mathrm{G}}(B)$ via the term in parentheses $[1 + \erf(z)]$ through (positive or negative) values of the parameter $\varsigma$.}.
While this formulation is elegant,
the piecewise definition in \Cref{eq:skewed-gaussian} has the pragmatic advantage
of being amenable to fast computation during fitting.
Specifically,
upon substituting \Cref{eq:skewed-gaussian} for $p(B)$ into \Cref{eq:polarization},
the solution to the integral can be written as~\cite{2008-Suter-M}:
\begin{equation}
	\label{eq:polarization-skewed-gaussian}
	P_{\mathrm{SG}}(t) = P_{\mathrm{SG}}^{-}(t) + P_{\mathrm{SG}}^{+}(t),
\end{equation}
where
\begin{widetext}
\begin{equation}
	\label{eq:polarization-skewed-gaussian-parts}
	P_{\mathrm{SG}}^{\pm}(t) = \left ( \frac{ \sigma_{\pm} }{\sigma_{+} + \sigma_{-}} \right ) \exp \left ( -\frac{\sigma_{\pm}^{2}t^{2} }{2} \right ) \left [ \cos ( \gamma_{\mu} B_{0} t + \phi ) \mp \erfi \left ( \frac{\sigma_{\pm} t}{\sqrt{2}} \right ) \sin ( \gamma_{\mu} B_{0} t + \phi ) \right ] ,
\end{equation}
\end{widetext}
and $\erfi (z)$ is the complex error function~\footnote{$\erfi (z)$ is usually defined in terms of one of several closely related functions (see e.g.,~\cite{2010-Olver-NISTHMF}). For example, our implementation~\cite{2008-Suter-M,2012-Suter-PP-30-69} used the confluent hypergeometric function of the first kind, $_{1}F_{1}(a, b, z)$, which was made available through the GNU Scientific Library~\cite{gsl} and a ``wrapper'' within the ROOT framework~\cite{1997-Brun-NIMA-389-81}.}.
We find that \Cref{eq:polarization-skewed-gaussian,eq:polarization-skewed-gaussian-parts}
give the best agreement with the signal in our \ch{Nb} samples over the full time range of
the measurement,
without overparamaterization~\footnote{A reasonable alternative to this could be to use a sum of $P_{\mathrm{G}}(t)$s; however, even with only two terms the sum's degrees of freedom would exceed that of \Cref{eq:polarization-skewed-gaussian,eq:polarization-skewed-gaussian-parts}.}.

Returning to our task of fitting the \gls{le-musr} data,
explicitly, we used the expression:
\begin{equation}
	\label{eq:fit-function}
	A(t) = A_{0} \left [ f P_{\mathrm{SG}}(t) + (1 - f) P_{\mathrm{G}}(t) \right ] ,
\end{equation}
where $A_{0}$ is an energy-dependent constant
(on the order of \num{\sim 0.2} here),
$f$ is the fraction of the signal originating from our sample
(typically \num{> 0.9}),
and the remaining terms $P_{\mathrm{SG}}(t)$ and $P_{\mathrm{G}}(t)$
were given by
\Cref{eq:polarization-skewed-gaussian,eq:polarization-skewed-gaussian-parts}
and
\Cref{eq:polarization-gaussian},
respectively.
Additionally,
all measurements for a given sample were fit simultaneously
(i.e., in a so-called ``global'' fit)
using a common $\phi$.
This constraint was necessary,
as the phase becomes ill-defined when $A(t)$ is strongly
damped and few full precession periods are resolved
(e.g., for measurements in Meissner state at high implantation energies,
where the $\mu^{+}$ stopping depths are far below the surface)~\footnote{A detailed account of how using a shared phase $\phi$ systematically affects the results when measuring Meissner screening profiles with \gls{le-musr} can be found elsewhere~\cite{2012-Hossain-PhD}, showing that the effect is minimal in all cases.}.
All fitting was performed using musrfit~\cite{2012-Suter-PP-30-69},
which makes use of the MINUIT2 minimization routines~\cite{2005-Hatlo-IEEETNS-52-2818}
implemented within the ROOT framework~\cite{1997-Brun-NIMA-389-81}.
In all cases,
this fitting approach yielded excellent agreement with the data
(reduced $\chi^{2} \approx 1.06$)
and a subset of the results are shown in \Cref{fig:baseline-spectra}.

\begin{figure}
	\centering
	\includegraphics[width=1.0\columnwidth]{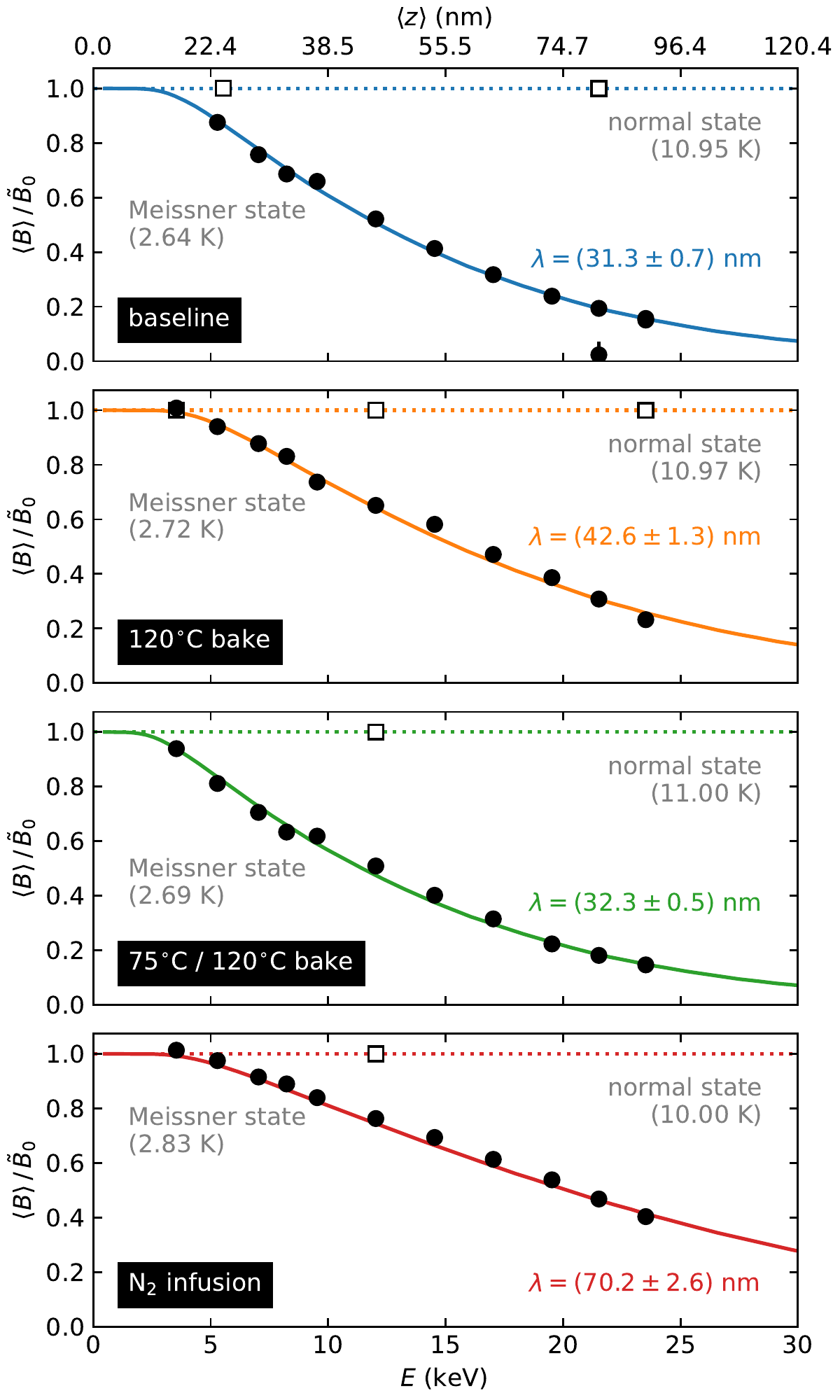}
	\caption{
		\label{fig:field-profiles}
		Plot of the mean magnetic field $\langle B \rangle$
		(normalized by the ``effective'' applied field $\tilde{B}_{0}$)
		sensed by the implanted $\mu^{+}$ at different energies $E$
		in \ch{Nb} samples that received different
		surface treatments
		(``baseline'',
		``\SI{120}{\celsius} bake''~\cite{2004-Ciovati-JAP-96-1591},
		``two-step bake''~\cite{arXiv:1806.09824},
		and
		``\ch{N2} infusion''~\cite{2017-Grassellino-SST-30-094004}
		---
		see \Cref{sec:experiment:samples}).
		For increasing $E$,
		the mean $\mu^{+}$ stopping depth $\langle z \rangle$ increases,
		covering a length scale comparable to the magnetic penetration
		depth $\lambda$.
		In the normal state ($T > T_{c}$),
		there is no depth dependence to $\langle B \rangle$ for any of the samples
		and its value corresponds to the applied magnetic field $B_{\mathrm{applied}}$.
		Conversely,
		in the Meissner state (\SI{\sim 2.7}{\kelvin}),
		$\langle B \rangle$ decays rapidly with increasing $E$
		above a threshold value,
		reflecting a small (non-superconducting) ``dead layer'' $d$ at the surface
		and
		the increased screening of $B_{\mathrm{applied}}$ at deeper depths.
		The solid and dashed colored lines represent a (global) fit of
		the data in both the normal and Meissner states using:
		\Cref{eq:london,eq:effective-field} to describe $B(z)$;
		\Cref{eq:stopping,eq:beta-pdf} to parameterize $\rho(z, E)$;
		and \Cref{eq:average-field} to convolve the terms into
		an expression for $\langle B \rangle (E)$
		(see \Cref{sec:results}).
		The fit quality is excellent,
		with the model capturing all features of the data.
		Values obtained for $\lambda$ are indicated in the inset
		of each subplot,
		while the full set of fit parameters are tabulated in \Cref{tab:results}.
	}
\end{figure}

In order to reconstruct the field profile below the surface,
at each implantation energy we identified the
mean field sensed by the implanted $\mu^{+}$ using~\cite{2008-Suter-M}:
\begin{equation}
	\label{eq:skewed-gaussian-mean}
	\langle B \rangle \equiv \int_{-\infty}^{+\infty} B \, p_{\mathrm{SG}}(B) \, \mathrm{d}B = B_{0} + \sqrt{\frac{2}{\pi}} \left ( \frac{ \sigma_{+} - \sigma_{-} }{ \gamma_{\mu} }\right ) ,
\end{equation}
and the results for each surface treatment are shown in
\Cref{fig:field-profiles}~\footnote{One can also use integral reconstruction to deduce $B(z)$ (see e.g.,~\cite{2004-Morenzoni-JPCM-16-S4583,2004-Suter-PRL-92-087001,2005-Suter-PRB-72-024506}); however, the approach relies on a \gls{fft} of the ``raw'' \gls{le-musr} data, making it numerically ill-posed.}.
As expected,
the $\langle B \rangle$s measured in the normal state are independent of implantation energy,
whereas $\langle B \rangle$ decreases monotonically with increasing $E$ in the Meissner state.
It is evident that the screening properties of each surface treatment are different;
the applied field is attenuated most strongly for the
``baseline''  and ``\SI{75}{\celsius}/\SI{120}{\celsius} bake''~\cite{arXiv:1806.09824} samples,
whereas the screening is weaker for the ``\SI{120}{\celsius} bake''~\cite{2004-Ciovati-JAP-96-1591}
and even more so for the ``\ch{N2} infusion''~\cite{2017-Grassellino-SST-30-094004} treatments.
Interestingly,
measurements in some of the samples at the lowest $E$s
show that $\langle B \rangle$ plateaus at a value close to the nominal applied field,
suggesting the presence of a thin layer near the surface where the external field isn't screened
(i.e., a so-called ``dead layer'' at the superconductor's surface).
Such a region is fairly generic and observed in a wide range of superconductors
(see e.g.,~\cite{2000-Jackson-PRL-84-4958,2004-Suter-PRL-92-087001,2005-Suter-PRB-72-024506,2014-Romanenko-APL-104-072601,2017-Junginger-SST-30-125013,2010-Kiefl-PRB-81-180502,2012-Ofer-PRB-85-060506,2013-Kozhevnikov-PRB-87-104508,2015-Stilp-PRB-89-020510,2018-Howald-PRB-97-094514}),
though there is considerable variability between materials or even samples
(e.g., as a result of surface roughness~\cite{2012-Lindstrom-PP-30-249,2014-Lindstrom-JEM-85-149,2016-Lindstrom-JSNM-29-1499}).

In order to evaluate the magnetic penetration depth $\lambda$,
it is necessary to construct a model capable of describing the data.
The model must account for two crucial details:
how the magnetic field is screened below the surface as a function of depth $z$;
and
the depth distribution $\rho(z, E)$ sampled by the implanted $\mu^{+}$.
We shall consider each of these below.
Note that while our approach differs somewhat from earlier work in \ch{Nb}
(see e.g.,~\cite{2005-Suter-PRB-72-024506,2014-Romanenko-APL-104-072601,2017-Junginger-SST-30-125013}),
it is capable of accurately reproducing all measured quantities
derived from our experiments.

First, we consider the magnetic field profile, $B(z)$,
below \ch{Nb}'s surface.
In the simplest case, $B(z)$ decreases exponentially with increasing depth, $z$,
in the Meissner state,
as predicted by the London model~\cite{1935-London-PRSLA-149-71}.
Recalling that our data suggests the presence of a ``dead layer'' at
the sample surface,
we incorporate this detail \emph{ad hoc} into the London result~\cite{1935-London-PRSLA-149-71}
with the expression (see e.g.,~\cite{2000-Jackson-PRL-84-4958,2010-Kiefl-PRB-81-180502}): 
\begin{equation}
	\label{eq:london}
	B(z) = \tilde{B}_{0} \times \begin{cases}
		1, & \text{for } z < d , \\
		\displaystyle \exp \left \{ -\frac{ ( z - d ) }{ \lambda } \right \} , & \text{for } z \geq d , 
	\end{cases}
\end{equation} 
where
$\lambda$ is the magnetic penetration depth,
$d$ is the thickness of the ``dead layer''
(i.e., where $\tilde{B}_{0}$ isn't screened),
and $\tilde{B}_{0}$ is the (effective) applied magnetic field.
The latter quantity is given by:
\begin{equation}
	\label{eq:effective-field}
	\tilde{B}_{0} = B_{\mathrm{applied}} \times \begin{cases}
		1, & \text{for } T > T_{c} , \\
		\left ( 1 - \tilde{N} \right )^{-1} , & \text{for } T \ll T_{c} , \\
	\end{cases}
\end{equation}
where $B_{\mathrm{applied}}$ is the applied magnetic field
and
$\tilde{N}$ is the sample's (effective) demagnetization factor~\cite{2018-Prozorov-PRA-10-014030}.
Note that the inclusion of the factor
$(1 - \tilde{N})^{-1}$ in \Cref{eq:effective-field}
accounts for any apparent ``enhancement'' of the applied field
due to the sample's geometry
(i.e., from flux expulsion in the Meissner state --- see e.g.,~\cite{2000-Brandt-PC-332-99}).
Though \Cref{eq:london} is rather simple
compared to other models for $B(z)$
(see e.g.,~\cite{1953-Pippard-PRSLA-216-547,1957-Bardeen-PR-108-1175}),
it sufficiently describes the behavior we observe (see below).

We now consider the $\mu^{+}$ implantation profiles.
As alluded to in \Cref{sec:experiment},
the slowing down of implanted $\mu^{+}$ is a stochastic process,
resulting in a \emph{distribution} of stopping depths
that can be reliably simulated~\cite{2002-Morenzoni-NIMB-192-245}
using Monte Carlo codes
such as TRIM.SP~\cite{1984-Eckstein-NIMB-2-550,1991-Eckstein-SSMS-10,1994-Eckstein-REDS-1-239}
(see \Cref{sec:trimsp} for specific details).
For our analysis,
it was convenient to have the ability to describe these profiles at
\emph{arbitrary} $E$, which was achieved by fitting the simulated
profiles
and interpolating their ``shape'' parameters.
We found that
the $\mu^{+}$ stopping probability, $\rho(z, E)$,
at a given $E$ can be described, in general, by:
\begin{equation}
	\label{eq:stopping}
	\rho(z, E) = \sum_{i}^{n} f_{i} p_{i} (z) ,
\end{equation}
where
$p_{i}(z)$ is a probability density function, 
$f_{i} \in [0, 1]$ is the $i^{\mathrm{th}}$ stopping fraction,
constrained such that
\begin{equation*}
	\sum_{i}^{n} f_{i} \equiv 1 ,
\end{equation*}
and $z$ is the depth below the surface.
For our target
(\ch{Nb2O5}(\SI{5}{\nano\meter})/\ch{Nb} --- see e.g.,~\cite{1987-Halbritter-APA-43-1}),
the stopping data are well-described using $n = 2$ and a $p(z)$ given by
a modified beta distribution~\cite{2004-Gupta-HBDA}.
Explicitly,
\begin{equation}
   \label{eq:beta-pdf}
   p (z) =
   \begin{cases}
         0, & \text{for } z < 0 , \\
         \dfrac{ \left ( z / z_{0} \right )^{\alpha -1}  \left (1 -  z / z_{0} \right )^{\beta - 1}  }{ z_{0} \, B ( \alpha, \beta ) } , & \text{for } 0 \leq z \leq z_{0} , \\
         0, & \text{for }  z > z_{0} ,
      \end{cases}
\end{equation}
where $z \in [0, z_{0}]$ is the depth below the surface
and $B ( \alpha, \beta )$ is the beta function:
\begin{equation*}
   B ( \alpha, \beta ) \equiv \frac{ \Gamma (\alpha ) \Gamma (\beta) }{ \Gamma ( \alpha + \beta ) } ,
\end{equation*}
with $\Gamma (s)$ denoting the gamma function:
\begin{equation*}
	% \label{eq:gamma}
	\Gamma (s) \equiv \int_{0}^{\infty} x^{s-1} \exp (-x) \, \mathrm{d}x .
\end{equation*}
Note that the ``extra'' $z_{0}$ in the denominator of \Cref{eq:beta-pdf}
ensures proper normalization of $p(z)$.

In order to achieve good ``coverage'' across the range of $E$s
achievable by \gls{le-musr}
(\SIrange{\sim 0.5}{\sim 30}{\kilo\electronvolt}),
we simulated $\mu^{+}$ stopping profiles in small energy
increments (\SI{500}{\electronvolt}) spanning the entire $E$-range.
We then fit each of the simulated stopping profiles using \Cref{eq:stopping,eq:beta-pdf}
and interpolated the resulting ``shape'' (i.e., fit)
parameters to generate $\rho(z, E)$ at arbitrary $E$.
Results from this procedure are shown in \Cref{fig:implantation-profiles},
in excellent agreement with the Monte Carlo simulations.

Following the above discussion,
with our expressions for
$B(z)$ [\Cref{eq:london,eq:effective-field}]
and
$\rho(z, E)$ [\Cref{eq:stopping,eq:beta-pdf}]
in hand,
it is now straightforward to construct an expression for
$\langle B \rangle$ that depends on $E$:
\begin{equation}
	\label{eq:average-field}
	\langle B \rangle (E) = \int_{0}^{\infty} B(z) \rho(z, E) \, \mathrm{d}z ,
\end{equation}
where the dependence on $E$ is accounted for \emph{implictly} by $\rho(z, E)$~\footnote{Formally, \Cref{eq:average-field} is the integral transform of $B(z)$ by the \emph{kernel} $\rho(z, E)$, wherein $B(z)$ is ``mapped'' from $z$-space to $\langle B \rangle (E)$ in $E$-space (see e.g.,~\cite{2005-Arfken-MMP-6}).}.
Note that, as described above,
$\rho(z, E)$'s ``shape'' parameters are all \emph{predetermined}
from fitting a series of implantation profiles and interpolating their values.
Consequently,
this approach uses the \emph{maximum} amount of
available information when fitting the data and does not,
for example,
assume that the average stopping depth, $\langle z \rangle$,
is an adequate proxy for the \emph{full} stopping distribution~\footnote{As noted elsewhere~\cite{2005-Suter-PRB-72-024506}, using $\langle z \rangle$ can influence the apparent ``curvature'' in the trend of $\langle B \rangle$, presumably because the mapping from $E$ to $\langle z \rangle$ is non-linear (see e.g., \Cref{fig:field-profiles}).}.
Therefore, \Cref{eq:average-field} depends on the main parameters that define the
shape of $B(z)$ [\Cref{eq:london,eq:effective-field}]:
$\lambda$, $d$, $B_{\mathrm{applied}}$, and $\tilde{N}$.
Before proceeding,
we point out that the integral \Cref{eq:average-field} must be evaluated
\emph{numerically};
however,
it was found that adaptive Gaussian quadrature routines
(see e.g.,~\cite{quadpack}),
which are widely available in free scientific software
(e.g., the Python package SciPy~\cite{2020-Virtanen-NM-17-261}),
are adequate for this task.
Fit results for each sample are given in \Cref{fig:field-profiles},
showing excellent agreement with the data,
and a tabulation of the resulting fit parameters is given \Cref{tab:results}.

\begin{table*}
	\centering
	\caption{
		\label{tab:results}
		Fit results for our \ch{Nb} samples with different surface treatments
		commonly used to fabricate \gls{srf} cavities (see \Cref{sec:experiment:samples}),
		obtained using the analysis approach described in \Cref{sec:results}
		(see also \Cref{fig:field-profiles}).
		Here,
		$T$ is the absolute temperature ,
		$B_{\mathrm{applied}}$ is the strength of the magnetic field
		applied parallel to the sample surface,
		$\tilde{N}$ is the sample's (effective) demagnetization factor,
		$d$ is the thickness of the (non-superconducting) ``dead layer'' at the
		sample surface,
		and $\lambda(T)$ is the magnetic penetration depth
		(measured at temperature $T$).
		Also included are quantities derived from \Cref{eq:two-fluid,eq:dirty,eq:effective-coherence-length}:
		the magnetic penetration depth at \SI{0}{\kelvin}, $\lambda_{0}$,
		the carrier mean-free-path, $\ell$,
		and
		the ``effective'' coherence length, $\xi_{0}^{\prime}$.
		For comparison,
		we have also included values for several \ch{Nb/Cu} films commonly used
		in \gls{srf} cavities~\cite{2017-Junginger-SST-30-125013}
		(obtained from a re-analysis of the data using the formalism described in \Cref{sec:results}),
		and
		results for a ``clean'' \ch{Nb/Al2O3} film~\cite{2005-Suter-PRB-72-024506}.
		The abbreviations listed with these samples correspond to:
		direct current magnetron sputtering (DCMS);
		high-power impulse magnetron sputtering (HIPIMS);
		and
		high-intensity and energy isotope mass separator on-line (HIE-ISOLDE).
		The dependence of $\lambda_{0}$ on $\ell$ is also shown in \Cref{fig:mean-free-path}.
	}
	% work around to get siunitx formatting in a ruledtabular-like environment 
	% https://tex.stackexchange.com/a/267904
	% https://tex.stackexchange.com/a/267916
	\footnotesize
	\begin{tabular*}{\textwidth}{l @{\extracolsep{\fill}} S S S S S S S S l}
	\botrule
	{Sample} & {$T$ (\si{\kelvin})} & {$B_{\mathrm{applied}}$ (\si{\milli\tesla})} & {$\tilde{N}$} & {$d$ (\si{\nano\meter})} & {$\lambda(T)$ (\si{\nano\meter})} & {$\lambda_{0}$ (\si{\nano\meter})} & {$\ell$ (\si{\nano\meter})} & {$\xi_{0}^{\prime}$ (\si{\nano\meter})} & Ref. \\
	\hline
	Nb (baseline) & 2.63 & 25.11 \pm 0.05 & 0.000 \pm 0.027 & 21.8 \pm 0.7 & 31.3 \pm 0.7 & 31.2 \pm 0.7 & 260 \pm 80 & 34.8 \pm 3.0 & This work \\
	Nb ($120^{\circ}$C bake) & 2.72 & 25.179 \pm 0.034 & 0.006 \pm 0.011 & 25.4 \pm 1.3 & 42.6 \pm 1.3 & 42.4 \pm 1.3 & 35 \pm 5 & 18.8 \pm 1.6 & This work \\
	Nb ($75^{\circ}$C / $120^{\circ}$C bake) & 2.69 & 25.162 \pm 0.032 & 0.000 \pm 0.028 & 18.7 \pm 0.6 & 32.3 \pm 0.5 & 32.2 \pm 0.5 & 175 \pm 34 & 32.8 \pm 2.6 & This work \\
	Nb (N$_{2}$ infusion) & 2.83 & 25.11 \pm 0.06 & 0.009 \pm 0.011 & 24.1 \pm 1.6 & 70.2 \pm 2.6 & 69.9 \pm 2.5 & 8.4 \pm 1.0 & 6.9 \pm 0.7 & This work \\
	\hline
	Nb/Cu (DCMS) & 3.25 & 15.14 \pm 0.05 & 0 & 14.3 \pm 1.2 & 51.1 \pm 1.2 & 50.7 \pm 1.2 & 19.6 \pm 2.2 & 13.2 \pm 1.1  & \cite{2017-Junginger-SST-30-125013} \\
	Nb/Cu (HIE-ISOLDE) & 2.65 & 15.010 \pm 0.024 & 0 & 13.1 \pm 0.5 & 37.8 \pm 0.7 & 37.7 \pm 0.7 & 59 \pm 7 & 23.9 \pm 1.7  & \cite{2017-Junginger-SST-30-125013} \\
	Nb/Cu (HIPIMS) & 3.75 & 15.09 \pm 0.04 & 0 & 17.3 \pm 0.6 & 34.6 \pm 0.4 & 34.1 \pm 0.4 & 106 \pm 14 & 29.2 \pm 2.1  & \cite{2017-Junginger-SST-30-125013} \\
	\hline
	\ch{Nb/Al2O3} (DCMS) &  & 8.82 & 0 & 2 \pm 2 &  & 27 \pm 3 & 359 & 36 & \cite{2005-Suter-PRB-72-024506} \\
	\botrule
\end{tabular*}
\end{table*}

\section{
	Discussion
	\label{sec:discussion}
}

From \Cref{fig:field-profiles},
it is clear that the different surface treatments affect the Meissner screening
profile in \ch{Nb} within the first \SI{\sim 150}{\nano\meter} below its surface.
As mentioned in \Cref{sec:results}, a hierarchy is evident;
the applied field is attenuated most strongly in the ``baseline'' sample,
yielding a $\lambda$ of \SI{31.3 \pm 0.7}{\nano\meter},
followed closely by the ``\SI{75}{\celsius}/\SI{120}{\celsius} bake'' treatment,
where $\lambda = \SI{32.3 \pm 0.5}{\nano\meter}$.
In the ``\SI{120}{\celsius} bake'' sample,
the screening was weakened significantly from the previous two treatments,
amounting to a magnetic penetration depth of $\SI{42.6 \pm 1.3}{\nano\meter}$.
This was diminished even further by the ``\ch{N2} infusion'' treatment, 
whose $\lambda = \SI{70.2 \pm 2.6}{\nano\meter}$.
The results suggest that further preparation beyond our ``baseline'' treatment
serves to diminish \ch{Nb}'s capacity to prevent magnetic flux from ``leaking''
below its surface in the Meissner state.

For a closer quantitative comparison between our results,
it is first necessary to account for the (minor) temperature differences
between measurements (see \Cref{fig:field-profiles}).
For this,
we used the well-known ``two-fluid'' expression~\cite{1996-Tinkham-IS-2}:
\begin{equation}
	\label{eq:two-fluid}
	\lambda (T) =  \frac{ \lambda_{0} }{ \sqrt{ 1 - \left ( T / T_{c} \right )^{4} } } ,
\end{equation}
where $\lambda_{0}$ is the magnetic penetration depth at \SI{0}{\kelvin},
to extrapolate the $\lambda$s down to absolute zero
(see \Cref{tab:results})~\footnote{Note that $T_{c}$ is essentially identical for all surface treatments used here (see e.g.,~\cite{2022-Turner-SR-12-5522}).}.
Extrapolating to this limit is convenient,
since at \SI{0}{\kelvin} we also have the simple relationship~\cite{1996-Tinkham-IS-2}:
\begin{equation}
	\label{eq:dirty}
	\lambda_{0} = \lambda_{L} \sqrt{ 1 + \frac{\xi_{0}}{\ell} } ,
\end{equation}
where $\lambda_{L}$ is the so-called London penetration depth,
$\xi_{0}$ is the Pippard~\cite{1953-Pippard-PRSLA-216-547} or \gls{bcs}~\cite{1957-Bardeen-PR-108-1175} coherence length,
and $\ell$ is the carrier mean-free-path
(i.e., the average distance traveled before being scattered).
As both $\lambda_{L}$ and $\xi_{0}$ can be regarded as material properties intrinsic to \ch{Nb},
differences in $\lambda_{0}$ can be understood in terms of different $\ell$s
for our samples.
By aid of \Cref{eq:dirty} and literature estimates~\footnote{For $\lambda_{L}$, we used a weighted average, correcting for temperature differences using \Cref{eq:two-fluid}. For $\xi_{0}$, we used a statistical average, as most studies do no quote uncertainties for their estimates.}
for both
$\lambda_{L} = \SI{29.01 \pm 0.10}{\nano\meter}$~\cite{1965-Maxfield-PR-139-A1515,1966-Finnemore-PR-149-231,1968-French-C-8-301,1973-Auer-PRB-7-136,1974-Varmazis-PRB-10-1885,1981-Epperlein-PBC-108-931,1984-Felcher-PRL-52-1539,1991-Weber-PRB-44-7585,1992-Korneev-PSPIE-1738-254,1994-Kim-JAP-75-8163,1995-Andreone-PRB-52-4473,1995-Zhang-PRB-52-10395,1998-Pronin-PRB-57-14416}
and
$\xi_{0} = \SI{40.3 \pm 3.5}{\nano\meter}$~\cite{1965-Maxfield-PR-139-A1515,1966-Finnemore-PR-149-231,1968-French-C-8-301,1973-Auer-PRB-7-136,1974-Varmazis-PRB-10-1885,1981-Donnelly-PVM-118,1981-Epperlein-PBC-108-931,1991-Weber-PRB-44-7585,1992-Wood-NIMA-314-86,1995-Andreone-PRB-52-4473,1998-Pronin-PRB-57-14416},
we calculate $\ell$ for our samples,
with the results tabulated in \Cref{tab:results}.
These values compare well with typical $\ell$s found in
\gls{srf} \ch{Nb}~\cite{2005-Casalbuoni-NIMA-538-45,2016-Martinello-APL-109-062601};
however,
to better understand their differences,
we must consider the material modifications introduced by these treatments.

We shall start with the ``baseline'',
which is simplest case to consider.
As described in \Cref{sec:experiment:samples},
this treatment first removes mechanical stresses through annealing and
afterwards purges surface imperfections in the topmost material
to mitigate any contamination from the furnace.
The procedure is highly successful,
as evidenced by our measured $\lambda$'s close proximity to $\lambda_{L}$,
suggesting that the level of impurities is low,
corresponding to an $\ell = \SI{260 \pm 80}{\nano\meter}$.
This is somewhat lower than the $\ell \sim \SI{810}{\nano\meter}$
expected for $\mathrm{RRR} \approx 300$ \ch{Nb}
(see e.g,~\cite{1968-Goodman-JPF-29-240,1972-Garwin-APL-20-154});
however,
we point out that our \emph{microscopic} method of determining $\ell$
only samples the spatial region probed by the $\mu^{+}$ beam,
making it more sensitive to the surface region where, for example,
interstitial impurities are likely more prevalent.
Similarly,
it was at first surprising to find that
non-local electrodynamics~\cite{1953-Pippard-PRSLA-216-547,1957-Bardeen-PR-108-1175}
were not necessary to describe the data;
however,
this is consistent with our $\ell$,
which equivalently yields a short ``effective'' coherence length $\xi_{0}^{\prime}$
(at \SI{0}{\kelvin})
according to~\cite{1996-Tinkham-IS-2}:
\begin{equation}
	\label{eq:effective-coherence-length}
	\frac{1}{ \xi_{0}^{\prime} } = \frac{1}{ \xi_{0} } + \frac{1}{\ell} .
\end{equation}
For the ``baseline'' sample,
we get $\xi_{0}^{\prime} = \SI{34.8 \pm 3.0}{\nano\meter}$,
which is very close to $\lambda_{0}$
and
equivalent to an $\xi_{0}^{\prime} / \lambda_{0} = \num{1.11 \pm 0.10}$.
Thus,
we conclude that this sample is close to the ``boundary'' where the influence of
non-local electrodynamics becomes significant.

We now consider the ``\SI{120}{\celsius} bake'' sample.
The main effect of the baking~\cite{2004-Ciovati-JAP-96-1591}
is to ``dissolve'' some of the
surface oxide into the bulk of \ch{Nb}.
This treatment instigated a refinement of the oxygen transport model in \ch{Nb}~\cite{2006-Ciovati-APL-89-022507}
which has received renewed attention as of late~\cite{2021-Lechner-APL-119-082601}.
Even before the invention of this ``recipe'',
oxygen diffusion profiles in \ch{Nb} were of interest for their
influence on the surface barrier associated with flux penetration~\cite{1978-vanderMey-PBC-95-369}.
Consistent with the empirical observation that this mild-baking
helps mitigate the so-called ``$Q$-slope'' observed in \gls{srf} cavities at high $E_{\mathrm{acc}}$,
we observe a $\lambda_{0}$ appreciably larger than $\lambda_{L}$
(equivalent to a reduced supercurrent density at the surface --- see e.g.,~\cite{2017-Kubo-SST-30-023001}),
accompanied by an
$\ell = \SI{35 \pm 5}{\nano\meter}$
and
$\xi_{0}^{\prime} = \SI{18.8 \pm 1.6}{\nano\meter}$.
These values are consistent with a sample whose surface region has been ``dirtied''
by the (intentional) addition of impurities.
Interestingly,
not only is this $\ell$ much larger than the values reported for this treatment
in another \gls{le-musr} study~\cite{2014-Romanenko-APL-104-072601},
the Meissner screening profile is also different.
While the bipartite behavior reported previously~\cite{2014-Romanenko-APL-104-072601}
has been suggested to originate from the baking~\cite{2004-Ciovati-JAP-96-1591}
producing an ``effective'' superconductor-superconductor bilayer~\cite{2014-Kubo-APL-104-032603,2017-Kubo-SST-30-023001,2019-Kubo-JJAP-58-088001}
(i.e., from a thin ``dirty'' surface on top of a ``clean'' bulk),
no evidence for such behavior is observed here.
In fact,
separate \gls{le-musr} measurements on \emph{real} bilayers~\cite{Asaduzzaman-tbp}
reveal screening profiles that are qualitatively distinct from those reported here
(see \Cref{fig:field-profiles}).
Thus,
we suggest that low-temperature baking~\cite{2004-Ciovati-JAP-96-1591}
does \emph{not} fundamentally alter the character of Meissner screening in \ch{Nb}
and that the earlier results~\cite{2014-Romanenko-APL-104-072601}
must find an alternative explanation.

Next, we consider the ``\SI{75}{\celsius}/\SI{120}{\celsius} bake'' sample.
Given what is known about mild baking~\cite{2004-Ciovati-JAP-96-1591},
the results for this two-step treatment~\cite{arXiv:1806.09824}
confound expectations.
The similarity of its $\lambda_{0}$ and derived quantities
to the ``baseline'' treatment
(see \Cref{tab:results})
suggests that the ``extra'' baking time
undermines the level of defects near the surface.
Explicit investigations into this matter are limited;
however,
one study using positron annihilation spectroscopy
proposed that the procedure ~\cite{2020-Wenskat-SR-10-8300}:
1) initially causes an increase in the \ch{Nb} vacancy concentration
through the decomposition of hydride-vacancy complexes;
2) that subsequent annealing at \SI{120}{\celsius} gradually removes the
complexes by thermally activated release;
and
3) that the remaining vacancies are vanquished by diffusion to trapping sites
and gradually annealed out.
While this mechanism is plausible,
it does not consider the dissolution of oxygen from the surface
during the second step~\cite{2004-Ciovati-JAP-96-1591},
which should have the \emph{opposite} influence on $\lambda$.
Thus,
we suggest that further investigation into
the near-surface chemical composition
(e.g., using secondary ion mass spectrometry)
is needed to be conclusive.

Lastly,
we consider the remaining surface treatment ``\ch{N2} infusion''~\cite{2017-Grassellino-SST-30-094004},
which is quite different from the other surface treatments.
In this ``recipe'',
\ch{N2} gas is intentionally introduced during baking to dope \ch{Nb} with nitrogen.
The ``infusion'' is performed at the relatively low temperature of \SI{120}{\celsius}
which limits the diffusivity of nitrogen~\cite{2020-Dhakal-PO-5-100034},
but mitigates the requirement of surface removal after the treatment
(cf.\ the original doping ``recipe''~\cite{2013-Grassellino-SST-26-102001}).
Given this treatment's substantial dopant ``supply''~\cite{2017-Grassellino-SST-30-094004}
and
nitrogen's diffusivity in \ch{Nb}
(see e.g.,~\cite{2020-Dhakal-PO-5-100034}),
it isn't surprising that we obtain our longest $\lambda_{0}$
of all the surface treatments,
and,
correspondingly,
the shortest $\ell$ and $\xi_{0}^{\prime}$
---
\SI{8.4 \pm 1.0}{\nano\meter}
and
\SI{6.9 \pm 0.7}{\nano\meter},
respectively.

\begin{figure}
	\centering
	\includegraphics[width=1.0\columnwidth]{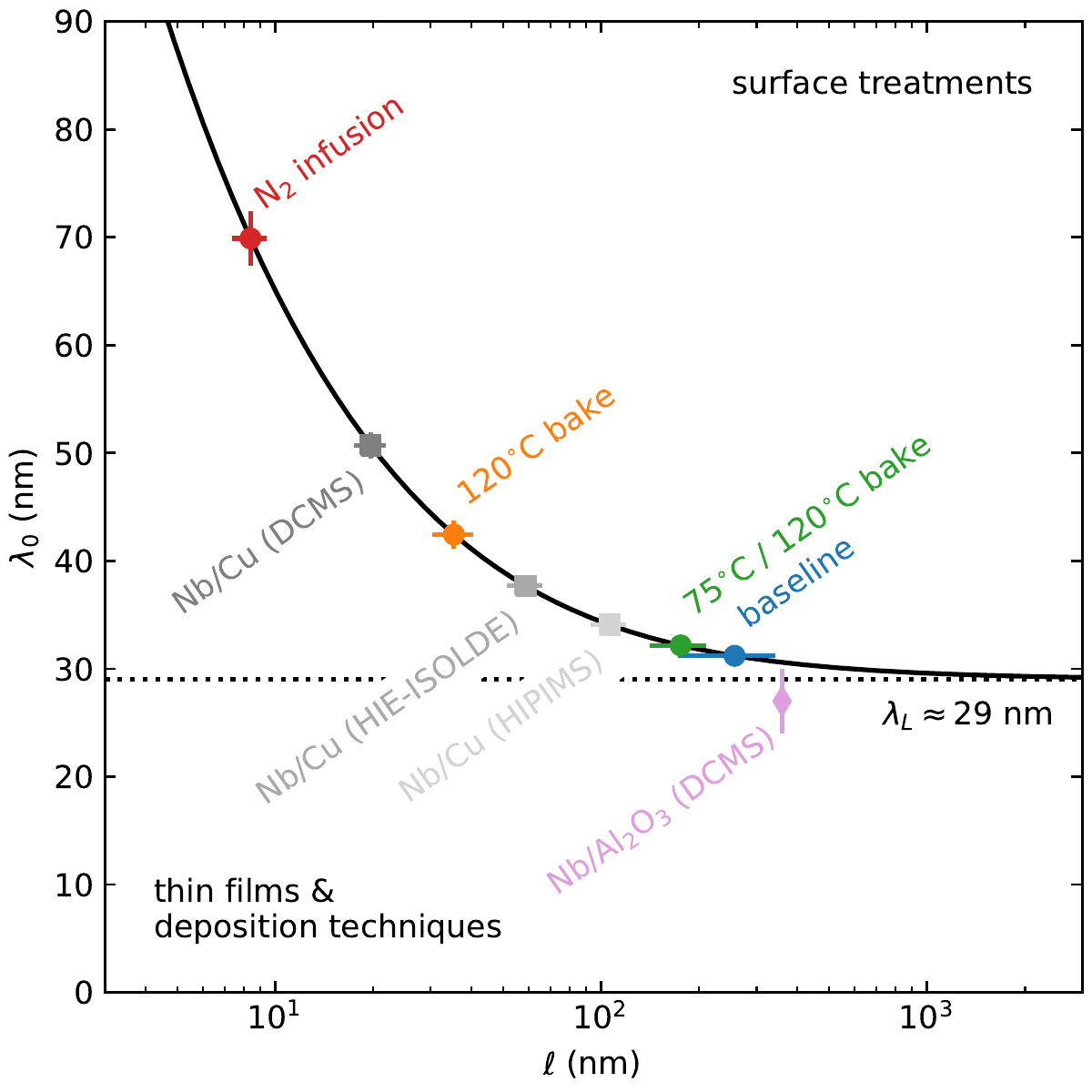}
	\caption{
		\label{fig:mean-free-path}
		Dependence of \ch{Nb}'s magnetic penetration depth at \SI{0}{\kelvin}, $\lambda_{0}$,
		on the carrier mean-free-path, $\ell$,
		for common surface treatments used to fabricate \gls{srf} cavities.
		The values were calculated according to \Cref{eq:dirty} (solid black line),
		using representative values for
		the London penetration depth
		$\lambda_{L} = \SI{29.01 \pm 0.10}{\nano\meter}$~\cite{1965-Maxfield-PR-139-A1515,1966-Finnemore-PR-149-231,1968-French-C-8-301,1973-Auer-PRB-7-136,1974-Varmazis-PRB-10-1885,1981-Epperlein-PBC-108-931,1984-Felcher-PRL-52-1539,1991-Weber-PRB-44-7585,1992-Korneev-PSPIE-1738-254,1994-Kim-JAP-75-8163,1995-Andreone-PRB-52-4473,1995-Zhang-PRB-52-10395,1998-Pronin-PRB-57-14416}
		and
		the \gls{bcs} coherence length
		$\xi_{0} = \SI{40.3 \pm 3.5}{\nano\meter}$~\cite{1965-Maxfield-PR-139-A1515,1966-Finnemore-PR-149-231,1968-French-C-8-301,1973-Auer-PRB-7-136,1974-Varmazis-PRB-10-1885,1981-Donnelly-PVM-118,1981-Epperlein-PBC-108-931,1991-Weber-PRB-44-7585,1992-Wood-NIMA-314-86,1995-Andreone-PRB-52-4473,1998-Pronin-PRB-57-14416}.
		Also included for comparison are values for \ch{Nb/Cu} films~\cite{2017-Junginger-SST-30-125013}
		prepared with different techniques
		(re-evaluated using the approach described in \Cref{sec:results}),
		and a very ``clean'' \ch{Nb/Al2O3} film~\cite{2005-Suter-PRB-72-024506}.
		All plotted values derived from this work are tabulated in \Cref{tab:results}.
	}
\end{figure}

Thus,
following the above discussion,
we propose that the observed hierarchy in $\lambda_{0}$ for the studied surface treatments
is readily explained by their propensity to dope \ch{Nb}'s near-surface region.
This (relatively light) doping alters $\ell$ in the spatial region sampled by \gls{le-musr},
resulting in $\lambda_{0} > \lambda_{L}$.
This relationship is shown graphically in \Cref{fig:mean-free-path},
accompanied by results from related studies for comparison~\cite{2005-Suter-PRB-72-024506,2017-Junginger-SST-30-125013}.
The results imply that either
$\ell$ is sufficiently homogeneous over the range of $\mu^{+}$ stopping depths
(see \Cref{fig:implantation-profiles})
to be encapsulated by a single (average) value
or
that the effect of any inhomogeneity in $\ell$
is beyond the detection limit of the current measurements.
Alternatively,
the largest inhomogeneity may be localized very close the surface,
comparable to the non-superconducting region observed 
of our samples (see \Cref{fig:field-profiles}),
considered below.

It is not uncommon to find a thin layer at a superconductor's surface that does not
screen an external field, colloquially called at ``dead layer''.
One typically accounts for this ``feature'' by incorporating the \emph{ad hoc} parameter $d$
into models of the screening profile --- see \Cref{eq:london}.
A literature survey suggests that $d$ is a generic feature of superconductors
(see e.g.,~\cite{2000-Jackson-PRL-84-4958,2004-Suter-PRL-92-087001,2005-Suter-PRB-72-024506,2010-Kiefl-PRB-81-180502,2012-Ofer-PRB-85-060506,2014-Romanenko-APL-104-072601,2013-Kozhevnikov-PRB-87-104508,2015-Stilp-PRB-89-020510,2017-Junginger-SST-30-125013,2018-Howald-PRB-97-094514}),
indicating that the quantity is representative of a particular \emph{sample},
rather than being intrinsic to the material.
For example, while a ``dead layer'' on the order of \SI{\sim 20}{\nano\meter}
is common for \gls{srf} \ch{Nb}~\cite{2014-Romanenko-APL-104-072601,2017-Junginger-SST-30-125013}
(which we also obtain here --- see \Cref{tab:results}),
values comparable to the thickness of the (native) surface oxide layer
(\SI{\sim 5}{\nano\meter}~\cite{1987-Halbritter-APA-43-1}) 
are found in high-quality epitaxial thin films~\cite{2005-Suter-PRB-72-024506,arXiv:2212.01137}.
Some of this variance is likely attributed to differences in surface roughness,
which can reduce a sample's screening capacity at the surface~\cite{2012-Lindstrom-PP-30-249,2014-Lindstrom-JEM-85-149,2016-Lindstrom-JSNM-29-1499};
however,
it alone cannot account for the full extent of $d$ in certain materials,
leading us to consider other possibilities.

Recently,
several authors have considered the possibility of $\lambda$ being
spatially \emph{inhomogeneous},
resulting from a varying defect concentration profile close to \ch{Nb}'s
surface~\cite{2019-Ngampruetikorn-PRR-1-012015,2020-Checchin-APL-117-032601}.
For a sufficiently high concentration of surface-localized defects,
it is plausible that $\lambda(z)$ could become long enough to qualitatively mimic
the effect of a ``dead layer''.
Such a scenario has already been considered theoretically for
$\lim_{z \to 0} \lambda(z) = \infty$~\cite{2014-Barash-JPCM-26-045702},
producing a gradual (rather than sharp) transition
between non-superconducting and superconducting regions.
While such an idea is intriguing,
our data in \Cref{fig:field-profiles} are not refined enough
to resolve such features and
further measurements
(e.g., with fine $E$ steps below \SI{\sim10}{\kilo\electronvolt})
are required to be more conclusive.
Such measurements may require a condensed \ch{N2} overlayer,
which can be grown \emph{in situ} at these low-$T$
(see e.g.,~\cite{2017-Junginger-SST-30-125013}).

Finally,
it is worth noting that a similar analysis approach to that described in
\Cref{sec:results} has also been employed in
\ch{^{8}Li} \gls{bnmr}~\cite{2015-MacFarlane-SSNMR-68-1,2022-MacFarlane-ZPC-236-757}
measurements on a \ch{Nb} thin film~\cite{arXiv:2212.01137}.
While the \gls{bnmr} technique shares many similarities with
and is complementary to \gls{le-musr}~\cite{2000-Kiefl-PB-289-640},
it has the advantage of being able to operate in (surface-parallel)
magnetic fields up to \SI{200}{\milli\tesla}~\cite{arXiv:2211.15619}, 
covering the operating conditions of \gls{srf} cavities
and close to \ch{Nb}'s $B_{\mathrm{sh}}$~\cite{2017-Junginger-SST-30-125012,2018-Junginger-PRAB-21-032002}.
Though equivalent measurements using \gls{le-musr} are not currently possible,
the results presented here will provide a good point of comparison
for future investigations using \gls{bnmr}.

\section{
	Conclusions
	\label{sec:conclusions}
}

Using \gls{le-musr},
we determined the Meissner screening profile in \ch{Nb} samples
that received surface treatments commonly used to prepare \gls{srf}
cavities.
In contrast to an earlier report~\cite{2014-Romanenko-APL-104-072601},
we find no evidence for any ``anomalous'' modifications to the Meissner profiles,
ruling out that low-temperature baking~\cite{2004-Ciovati-JAP-96-1591},
two-step baking~\cite{arXiv:1806.09824},
or \ch{N2} infusion~\cite{2017-Grassellino-SST-30-094004}
produces an ``effective'' bilayer superconductor~\cite{2014-Kubo-APL-104-032603,2017-Kubo-SST-30-023001,2019-Kubo-JJAP-58-088001}.
Instead,
we find that the observed field screening is well-described
by a simple London model~\cite{1935-London-PRSLA-149-71},
with magnetic penetration depths
(extrapolated to \SI{0}{\kelvin}) of:
\SI{31.3 \pm 0.7}{\nano\meter} for the ``baseline'' sample;
\SI{42.6 \pm 1.3}{\nano\meter} for the ``\SI{120}{\celsius} bake'' treatment;
\SI{32.3 \pm 0.5}{\nano\meter} for the ``\SI{75}{\celsius}/\SI{120}{\celsius} bake'' recipe;
and
\SI{70.2 \pm 2.6}{\nano\meter} for ``\ch{N2} infusion''.
Differences in screening properties between surface treatments can be
explained by changes to the carrier mean-free-paths resulting from 
dopant profiles near \ch{Nb}'s surface.
A relatively large (\SI{\sim 20}{\nano\meter}) non-superconducting ``dead layer''
was found in all samples, exceeding the thickness of the native oxide layer that forms
at \ch{Nb}'s surface~\cite{1987-Halbritter-APA-43-1}.
This observation may suggest a narrow region near the surface where $\lambda$
is \emph{depth-dependent}~\cite{2014-Barash-JPCM-26-045702,2019-Ngampruetikorn-PRR-1-012015,2021-Lechner-APL-119-082601}.
Further \gls{le-musr} experiments,
with finer steps energy steps where $E < \SI{10}{\kilo\electronvolt}$
may illuminate the matter.

\begin{acknowledgments}
	We thank:
	P.~Kolb, R.~E.~Laxdal, W.~A.~MacFarlane, and E.~Thoeng for useful discussions;
	TRIUMF's \gls{srf} group for providing several of the \ch{Nb} samples
	(``baseline'', ``\SI{120}{\celsius} bake'', and \SI{75}{\celsius}/\SI{120}{\celsius} bake'');
	and
	M.~Martinello for preparing the ``\ch{N2} infusion'' sample.
	This work is based on experiments performed at the Swiss Muon Source S$\mu$S,
	Paul Scherrer Institute, Villigen, Switzerland.
	Financial support was provided by an \gls{nserc} Award to T.~Junginger.
\end{acknowledgments}

\appendix

\section{
	Simulating Muon Implantation
	\label{sec:trimsp}
}

As is clear from \Cref{sec:results},
a crucial aspect of our analysis is the inclusion of simulated $\mu^{+}$
stopping profiles,
which serve as the kernel of the integral transform defined
in \Cref{eq:average-field}.
Given their importance,
we now consider some of their details further.

In general,  
energetic charged particles lose their kinetic energy through the interaction
with matter encountered along their trajectory.
For a particle penetrating a (solid) target,
this happens through a series collisions with the host's electrons and nuclei~\footnote{Radiative processes (e.g., brehmsstrahlung) are also important at high projectile energies; however, for the energies used in \gls{le-musr}, they are unimportant and we shall not consider them further.}.
In this slowing process,
the average energy loss per unit distance is often called the stopping power:
\begin{equation}
	\label{eq:stopping-power}
	S \equiv -\frac{\mathrm{d}E}{\mathrm{d}z} ,
\end{equation}
where $E$ is the energy and $z$ the position of the projectile.
Naturally, $S$ can be thought of as a property specific to particular
projectile/target combination.
Typically, $S$ is decomposed into electronic, $S_{e}$, and nuclear, $S_{n}$,
contributions:
\begin{equation}
	S = S_{n} + S_{e} ,
\end{equation}
such that their contributions may be treated separately (see below).
Practically, it is convenient to \emph{normalize} each $S_{i}$ by the target's
number density, $n$, converting them into stopping \emph{cross sections}:
\begin{equation}
	\label{eq:stopping-cross-section}
	\tilde{S}_{i} \equiv \frac{S_{i}}{n} .
\end{equation}
Note that $\tilde{S}_{i}$ is typically reported in ``odd'' looking units
(e.g., \SI{e-15}{\electronvolt\centi\meter\squared\per\atom}).

As there are an endless number of projectile/target combinations,
significant effort has been invested in predicting $\tilde{S}$,
which can be used to calculate a projectile's \emph{range}
(i.e., its average implantation depth) in \emph{any} target via:
\begin{equation}
	\label{eq:range}
	\langle z \rangle = \int_{E}^{0} \frac{1}{ n \tilde{S} } \, \mathrm{d}E .
\end{equation}
This task, however, is formidable and no single theoretical formalism
dominates to-date.
Though progress with theory continues to be made
(see e.g.,~\cite{2019-Schinner-NIMB-460-19}),  
a phenomenological treatment of $\tilde{S}_{i}$ is often used,
wherein measured values are parameterized using a semi-empirical model.
At the implantation energies used in a \gls{le-musr} experiment~\cite{2004-Morenzoni-JPCM-16-S4583,2004-Bakule-CP-45-203},
the \emph{electronic} contribution to $\tilde{S}$ has the greatest impact on a muon's
range and we now focus on $\tilde{S}_{e}$.

One parameterization of $\tilde{S}_{e}$ that has been used for over half
a century are the so-called Varelas-Biersack formulas~\cite{1970-Varelas-NIM-79-213},
which describe the stopping cross section in terms of five
coefficients, $A_{i}$, within the low-energy implantation regime
(\SIrange{1}{1000}{\kilo\electronvolt}).
In this treatment,
the energy dependence is expressed in terms of the scaled quantity:
\begin{equation}
	\label{eq:reduced-energy}
	\tilde{E} \equiv E \left ( \frac{ \mathrm{u} }{ m } \right ) ,
\end{equation}
where $E$ is the projectile energy (in \si{\kilo\electronvolt})
and
$m$ is its mass (in \si{\amu}~\cite{2021-Tiesinga-RMP-93-025010})~\footnote{The inclusion of the \si{\amu}~\cite{2021-Tiesinga-RMP-93-025010} in \Cref{eq:reduced-energy} keeps the ``scaling'' factor dimensionless~\cite{1993-ICRU-49}; however, the approach is not adopted by all authors~\cite{1977-Anderson-SRIM-3}.}.
For muons (i.e., light protons),
$\tilde{S}_{e}$ is given by~\cite{1970-Varelas-NIM-79-213,1977-Anderson-SRIM-3,1993-ICRU-49}:
\begin{equation}
	\label{eq:varelas-biersack}
	\tilde{S}_{e} = \begin{cases} 
		 A_{1} \sqrt{\tilde{E}} , & \SI{1}{\kilo\electronvolt} \leq \tilde{E} < \SI{10}{\kilo\electronvolt}, \\
		\dfrac{ s_{\mathrm{low}}(\tilde{E}) \, s_{\mathrm{high}}(\tilde{E}) }{ s_{\mathrm{low}}(\tilde{E}) + s_{\mathrm{high}}(\tilde{E}) } , & \SI{10}{\kilo\electronvolt} \leq \tilde{E} < \SI{1}{\mega\electronvolt}, \\
   \end{cases}
\end{equation}
where
\begin{equation}
	\label{eq:s-low}
	s_{\mathrm{low}}(\tilde{E}) = A_{2} \tilde{E}^{0.45} ,
\end{equation}
and 
\begin{equation}
	\label{eq:s-high}
	s_{\mathrm{high}}(\tilde{E}) = \left ( \frac{ A_{3} }{ \tilde{E} } \right ) \ln \left ( 1 + \frac{A_{4}}{\tilde{E}} + A_{5} \tilde{E} \right ) .
\end{equation}
While \Cref{eq:varelas-biersack,eq:s-low,eq:s-high} require five $A_{i}$s,
smooth continuity implies that:
\begin{equation*}
	A_{1} \equiv \left ( \frac{1}{\sqrt{ \SI{10}{\kilo\electronvolt} }} \right ) \frac{ s_{\mathrm{low}}( \SI{10}{\kilo\electronvolt} ) \, s_{\mathrm{high}}( \SI{10}{\kilo\electronvolt} ) }{ s_{\mathrm{low}}( \SI{10}{\kilo\electronvolt} ) + s_{\mathrm{high}}( \SI{10}{\kilo\electronvolt} ) } ,
\end{equation*}
reducing the number of ``free'' parameters from five to four.
In fact, \Cref{eq:varelas-biersack,eq:s-low,eq:s-high} have proved to be so
useful that several compilations of the $A_{i}$s have been made~\cite{1977-Anderson-SRIM-3,1993-ICRU-49}
for the purpose for facilitating simulations of ion-implantation.
Important for us,
the Monte Carlo code TRIM.SP~\cite{1984-Eckstein-NIMB-2-550,1991-Eckstein-SSMS-10,1994-Eckstein-REDS-1-239}
relies on these tabulated values;
however, the most recent compilation~\cite{1993-ICRU-49} is nearly
\SI{\sim 30}{years} old
and a major concern is its current validity.

\begin{figure}
	\centering
	\includegraphics[width=1.0\columnwidth]{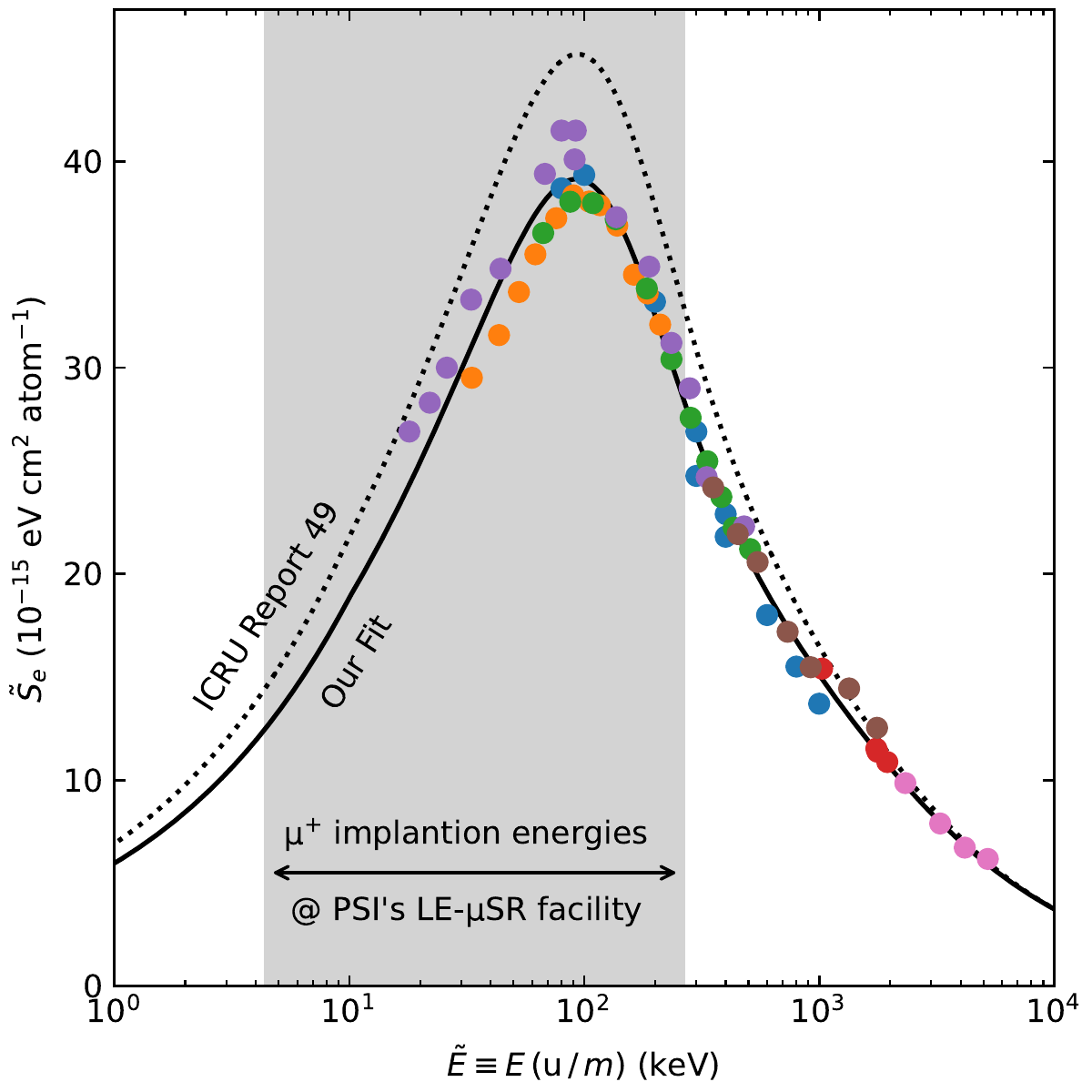}
	\caption{
		\label{fig:stopping-Nb}
		Electronic stopping cross section, $\tilde{S}_{e}$,
		as a function of scaled energy, $\tilde{E}$,
		for \ch{H} isotopes implanted in \ch{Nb}.
		The filled circles are measured values, with colors denoting results
		from different studies~\cite{1984-Sirotinin-NIMB-4-337, 1986-Bauer-NIMB-13-201, 1988-Ogino-NIMB-33-155, 2020-Moro-PRA-102-022808}.
		Note that results from one early report~\cite{1973-Behr-TSF-19-247},
		which are clear outliers, have been omitted.
		Different parameterizations of the cross sections using
		\Cref{eq:varelas-biersack,eq:s-low,eq:s-high}
		(i.e., the Varelas-Biersack formula~\cite{1970-Varelas-NIM-79-213})
		are also given,
		with our fit giving the best agreement with \emph{all} (reliable) experimental data,
		improving on the older tabulated values~\cite{1977-Anderson-SRIM-3,1993-ICRU-49}.
		The $A_{i}$s determined from our fit are listed in the \Cref{tab:varelas-biersack-params}.
		The experimental data was curated by the \gls{iaea}~\cite{2017-Montanari-NIMB-408-50}.
	}
\end{figure}

\begin{table*}
	\caption{
		\label{tab:varelas-biersack-params}
		Values for the $A_{i}$s in the (empirical) Varelas-Biersack formula~\cite{1970-Varelas-NIM-79-213}
		--- \Cref{eq:varelas-biersack,eq:s-low,eq:s-high} ---
		parameterizing the electronic stopping cross section, $\tilde{S}_{e}$,
		for \ch{H} isotopes implanted in \ch{Nb}.
		Plots of \Cref{eq:varelas-biersack,eq:s-low,eq:s-high} using the two sets of coefficients,
		along with all (reliable) experimental data~\cite{1984-Sirotinin-NIMB-4-337, 1986-Bauer-NIMB-13-201, 1988-Ogino-NIMB-33-155, 2020-Moro-PRA-102-022808}
		from the \gls{iaea} database~\cite{2017-Montanari-NIMB-408-50}
		are shown in \Cref{fig:stopping-Nb}.
		Note that, for brevity, we have used $\si{\stoppingunit} \equiv \SI{e-15}{\electronvolt\centi\meter\squared\per\atom}$
		when expressing the units of some of the $A_{i}$s.
	}
\begin{tabular*}{\textwidth}{S @{\extracolsep{\fill}} S S S S l}
	\botrule
	{$A_{1}$ (\si{\stoppingunit\kilo\electronvolt\tothe{-1/2}})} & {$A_{2}$ (\si{\stoppingunit\kilo\electronvolt\tothe{-0.45}})} & {$A_{3}$ (\SI{e4}{\stoppingunit\kilo\electronvolt})} & {$A_{4}$ (\SI{e2}{\kilo\electronvolt})} & {$A_{5}$ (\SI{e-3}{\per\kilo\electronvolt})} & {Ref.} \\ 
	\hline
	5.96 \pm 0.10 & 6.73 \pm 0.08 & 1.03 \pm 0.13 & 2.8 \pm 0.5 & 3.9 \pm 0.9 & This work \\
	6.901         & 7.791         & 0.9333        & 4.427       & 5.587       & \cite{1977-Anderson-SRIM-3,1993-ICRU-49} \\
	\botrule
\end{tabular*}
\end{table*}

Quite recently, a large database of electronic stopping cross sections
(originally compiled by the late H.~Paul)
has been released by the \gls{iaea}~\cite{2017-Montanari-NIMB-408-50},
greatly facilitating the comparison of experimental data with different models.
Using the compilation,
we compared different parameterizations of $\tilde{S}_{e}$ for \ch{H}-isotopes
in \ch{Nb} against available experimental
data~\cite{1984-Sirotinin-NIMB-4-337, 1986-Bauer-NIMB-13-201, 1988-Ogino-NIMB-33-155, 2020-Moro-PRA-102-022808}
(omitting clear outliers~\cite{1973-Behr-TSF-19-247}),
as shown in \Cref{fig:stopping-Nb}.
We found that the earlier tabulations~\cite{1977-Anderson-SRIM-3,1993-ICRU-49}
\emph{overestimate} $\tilde{S}_{e}$, likely due to lack of available data
at their time of publication~\footnote{Consequently, it is likely that the $\mu^{+}$ range in \ch{Nb} (or \ch{Nb} layers) was \emph{underestimated} in earlier simulations (see e.g.,~\cite{2005-Suter-PRB-72-024506,2014-Flokstra-PRB-89-054510,2014-Romanenko-APL-104-072601,2015-DiBernardo-PRX-5-041021,2016-Flokstra-NP-12-57,2017-Junginger-SST-30-125013,2018-Flokstra-PRL-120-247001,2019-Flokstra-APL-115-072602,2019-Stewart-PRB-100-020505,2020-Krieger-PRL-125-026802,2021-Rogers-CP-4-69,2021-Flokstra-RPB-104-L060506,2021-Alpern-PRM-5-114801}).}.
In contrast,
our fit to the Varelas-Biersack formulas~\cite{1970-Varelas-NIM-79-213}
[\Cref{eq:varelas-biersack,eq:s-low,eq:s-high}]
gives the best agreement with \emph{all} (reliable) experimental data,
and we used our new $A_{i}$s (listed in \Cref{tab:varelas-biersack-params})
in all TRIM.SP~\cite{1984-Eckstein-NIMB-2-550,1991-Eckstein-SSMS-10,1994-Eckstein-REDS-1-239}
simulations of $\mu^{+}$ implantation~\footnote{A similar check was also performed for \ch{H}-isotopes implanted in \ch{O2}; however, no meaningful deviation from the earlier tabulations~\cite{1977-Anderson-SRIM-3,1993-ICRU-49} was found.}.

Finally,
we conclude this section with some explicit details of the Monte Carlo
simulations of $\mu^{+}$ implantation in our target,
\ch{Nb2O5}(\SI{5}{\nano\meter})/\ch{Nb},
using TRIM.SP~\cite{1984-Eckstein-NIMB-2-550,1991-Eckstein-SSMS-10,1994-Eckstein-REDS-1-239}.
In these simulations, the projectile's trajectory is calculated step-by-step
with the assumption that its direction changes with each binary (nuclear)
collision and that its path remains straight while in ``free flight''.
Energy is lost via the (inelastic) electronic contribution to stopping,
which is treated independently from the nuclear contribution.
The projectile is said to have ``come to rest'' once its energy drops below a
threshold, whereafter its final position is recorded in a histogram,
which is output at the end of the simulation.
The choices below mainly originate from earlier work that
systematically compared $\mu^{+}$ stopping results against \gls{le-musr}
data~\cite{2002-Morenzoni-NIMB-192-245}.

Implantation profiles were simulated at select energies, $E$,
between \SIrange{0.5}{30}{\kilo\electronvolt}.
Each simulation used \num{e5} projectiles,
whose exact implantation energies were assumed to follow a normal distribution centered at $E$
with a width of
\SI{450}{\electronvolt}~\footnote{The finite width of the distribution is used to account for the spread in energies of the $\mu^{+}$ eluting from the cryocrystal moderator~\cite{2001-Prokscha-ASS-172-235} and the staggling introduced from passage through the carbon foil detector~\cite{2015-Khaw-JI-10-P10025}.}.
Any individual projectiles with an $E \leq \SI{0}{\electronvolt}$ were discarded.
The projectile angle of incidence (relative to the surface normal)
followed a normal distribution centered at \SI{0}{\degree} with a width of \SI{15}{\degree}.
A hydrogen-like projectile was assumed, but with the muon's mass,
$m_{\mu} = \SI{0.113 428 9259 \pm 0.000 000 0025}{\amu}$~\cite{2021-Tiesinga-RMP-93-025010},
which is about $\sim 1 / 9$\textsuperscript{th} the mass of a proton,
$m_{p} = \SI{1.007 276 466 621 \pm 0.000 000 000 053}{\amu} $~\cite{2021-Tiesinga-RMP-93-025010}.
Interactions between the projectile and target atoms were treated using a Moli\`{e}re-type
screened Coulomb potential~\cite{1947-Moliere-ZNA-3-133}
with a Firsov screening length~\cite{1958-Firsov-SPJETP-6-534}.
Interactions between different target atoms were treated using the so-called
\ch{Kr-C} potential~\cite{1977-Wilson-PRB-15-2458,1991-Eckstein-SSMS-10}.
The inelastic energy loss of the $\mu^{+}$ projectiles was treated using
the Varelas-Biersack model~\cite{1970-Varelas-NIM-79-213}
[\Cref{eq:varelas-biersack,eq:s-low,eq:s-high}]
and either tabulated~\cite{1977-Anderson-SRIM-3,1993-ICRU-49}
or re-derived stopping cross sections (see \Cref{fig:stopping-Nb,tab:varelas-biersack-params}).
For the target atoms,
this was done using an equipartition of
Oen-Robinson (local)~\cite{1976-Oen-NIM-132-647}
and Lindhard-Scharff (non-local)~\cite{1961-Lindhard-PR-124-128,1963-Lindhard-MFMDVS-33-14}
models.
A cutoff energy of \SI{0.5}{\electronvolt} was chosen for the projectiles.
Further explanation can be found elsewhere
(see e.g.,~\cite{1991-Eckstein-SSMS-10}).
To calculate the stopping power of \ch{Nb2O5},
the so-called Bragg rule was used
without any additional ``compound'' corrections
(see e.g.,~\cite{1991-Eckstein-SSMS-10}).
To speed up the simulations, sputtering effects were omitted
(i.e., no recoils were generated).
Typical stopping profiles produced from the simulations are shown in
\Cref{fig:implantation-profiles}.

% \bibliography{references.bib,unpublished.bib}


\end{document}